\documentclass[a4paper,11pt]{article}
\pdfoutput=1
\usepackage[pdftex]{graphics}
\usepackage{jheppub}
\usepackage{graphicx}
\usepackage{epsfig}
\usepackage{amsmath, amssymb}
\usepackage{dcolumn}
\usepackage{bm}
\usepackage{amsfonts}

\usepackage{amssymb}
\usepackage{subfig}

\def\dj{\hbox{d\kern-0.347em \vrule width 0.3em height 1.252ex depth
-1.21ex \kern 0.051em}}

\newcommand{\be}{\begin{equation}}
\newcommand{\ee}{\end{equation}}
\newcommand{\ben}{\begin{equation*}}
\newcommand{\een}{\end{equation*}}
\newcommand{\bea}{\begin{eqnarray}}
\newcommand{\eea}{\end{eqnarray}}
\newcommand{\bean}{\begin{eqnarray*}}
\newcommand{\eean}{\end{eqnarray*}}
\newcommand{\brr}{\begin{array}}
\newcommand{\err}{\end{array}}
\newcommand{\bc}{\begin{center}}
\newcommand{\ec}{\end{center}}

\newcommand{\gsim}{\,\raisebox{-0.6ex}{$\buildrel > \over \sim$}\,}

\newcommand{\OO}{{\cal O}}

\newcommand{\eps}{\epsilon}

\newcommand{\Om}{\Omega}

\newcommand{\OM}{\Omega_{\rm m,0}}

\newcommand{\ODE}{\Omega_{\rm de,0}}

\newcommand{\lzr}{(z)}
\newcommand{\hyper}{\ {}_2\!F_1}

\title{A parametrization of the growth index of matter perturbations in various Dark Energy models and observational prospects using a Euclid-like survey}

\author[a]{Alicia Bueno Belloso,}
\author[a,b]{Juan Garc\'{\i}a-Bellido,}
\author[a]{and Domenico Sapone}

\affiliation[a]{Instituto de F\'\i sica Te\'orica, Universidad Aut\'onoma de Madrid, Cantoblanco 28049 Madrid, Spain}
\affiliation[b]{Institute de Physique Th\'eorique, Universit\'e de Gen\`eve, 24 quai E. Ansermet, 
1211 Gen\`eve 4, Switzerland}

\emailAdd{alicia.bueno@uam.es}
\emailAdd{juan.garciabellido@uam.es}
\emailAdd{domenico.sapone@uam.es}

\abstract{
We provide exact solutions to the cosmological matter perturbation equation in a homogeneous FLRW universe with a vacuum energy that can be parametrized by a constant equation of state parameter $w$ and a very accurate approximation for the Ansatz $w(a)=w_0+w_a(1-a)$. We compute the growth index $\gamma=\log f(a)/\log\Om_m(a)$, and its redshift dependence, using the exact and approximate solutions in terms of Legendre polynomials and show that it can be parametrized as $\gamma(a)=\gamma_0+\gamma_a(1-a)$ in most cases. We then compare four different types of dark energy (DE) models: $w\Lambda$CDM, DGP, $f(R)$ and a LTB-large-void model, which have very different behaviors at $z\gsim1$. This allows us to study the possibility to differentiate between different DE alternatives using wide and deep surveys like Euclid, which will measure both photometric and spectroscopic redshifts for several hundreds of millions of galaxies up to redshift $z\simeq 2$. We do a Fisher matrix analysis for the prospects of differentiating among the different DE models in terms of the growth index, taken as a given function of redshift or with a principal component analysis, with a value for each redshift bin for a Euclid-like survey. We use as observables the complete and marginalized power spectrum of galaxies $P(k)$ and the Weak Lensing (WL) power spectrum. We find that, using $P(k)$, one can reach (2\%, 5\%) errors in $(w_0, w_a)$, and (4\%, 12\%) errors in $(\gamma_0, \gamma_a)$, while using WL we get errors at least twice as large. These estimates allow us to differentiate easily between DGP, $f(R)$ models and $\Lambda$CDM, while it would be more difficult to distinguish the latter from a variable equation of state parameter or LTB models using only the growth index.}

\keywords{Cosmological perturbations, General Relativity, Dark Energy\\[2mm]
{\sf Preprint:} {\tt IFT-UAM/CSIC-11-24}}

\begin{document}

\maketitle

\section{Introduction}\label{introduction}

With the turn of the millenium we have entered a new era in which cosmological observations have improved to the level that we can start to define a Standard Model of Cosmology based on the CDM paradigm plus some sort of vacuum energy responsible for the observed dimming of distant supernovae. The nature of either Dark Matter (DM) or Dark Energy (DE) remains a mystery, in spite of the improved determinations of their contribution to the total energy density of the Universe. 

While the nature of Dark Matter seems less uncertain (most cosmologists are in favor of a particle physics origin), that of Dark Energy is still unexplored territory. In the last decade there has been a plethora of proposals to account for the observed acceleration of the universe. All these proposals fall into four main categories: i) the inclusion of some extra field (scalar, vector or tensor), coupled or not to the rest of matter, like in quintessence, chameleon, vector dark energy or massive gravity; ii) the extension of GR by inclusion of higher order terms in the Einsten-Hilbert action, like f(R) theories, Gauss-Bonnet terms, etc.; iii) the modification of gravity on large scales by introduction of extra dimensions, like in the Dvali-Gabadadze-Porrati model, Kaluza-Klein gravity, etc.; iv) the reinterpretation in terms of a nontrivial spatial geometry, like in large-void inhomogeneous LTB models. For a recent review on DE modeling, see for example~\cite{tsujikawa,saponeDE}.

All of these proposals have very specific predictions for the background evolution of cosmological space-times, and most of them can be well fitted to the present observations with just a few phenomenological parameters: the equation of state, the speed of sound, the coupling between DM and DE, bulk viscosity, etc. However, in order to discriminate between the different alternatives it has been realized that one has to go beyond the background evolution and start to consider also the theory of linear cosmological perturbations and parametrize their evolution in terms of the growth function and growth index, as well as the shift parameter.

At the moment, the main observables used to constrain the dark sector are the cepheids and supernovae magnitudes for the determination of the expansion rate as a function of redshift, see~\cite{riess} for the most recent measurements; the power spectrum of matter and the baryon acoustic oscillation (BAO) scale for the determination of the matter content as a function of redshift, see~\cite{percivalbao,percivalpk}, together with the cosmic microwave background for the determination of global spatial curvature and the asymptotic values of cosmological parameters~\cite{komatsu}, and the weak lensing shear power spectrum~\cite{refregier}, or the ISW-galaxy cross-correlation~\cite{taburet}, for consistency of the whole scenario. Most of these measurements are rather preliminary and suggest a detection of DE at the 2 to 3 sigma level. However, a significant improvement is expected in the present decade thanks to Planck~\cite{planck,planckbluebook}, DES~\cite{des}, BOSS~\cite{boss} and, in the future, the Euclid survey~\cite{euclid}.

In this paper, we study the prospects that a survey like Euclid would have in distinguishing between the four main classes of DE models. In the process, we find exact and approximate solutions for the growth index in terms of simple functions, for $\Lambda$CDM models with a constant and variable equation of state parameter. We then propose a simple parametrization of the growth index that fits well the recent history, except for extreme models like $f(R)$. We note that while the background parameters $H(z)$, $\Omega_M(z)$ and $w(z)$ seem to be rather similar to those of $\Lambda$CDM, the growth index can differ significantly for most classes of models. 

The parametrization of the density contrast with the growth index was first introduced by Peebles in 1980 \cite{peebles}. We knew that the rate of growth of structures should be a function of the matter density. Several parametrization attempts including a power law expansion or simply the square root of the matter density parameter did not quite fit the data. Finally, the $\gamma$ parametrization was the most widely accepted one; however, we will have to wait for a direct measurements of the growth index, which will only be possible with the next generation of experiments. It is therefore tantalizing to explore the possibilities of distinguishing between different DE models with a better determination of the growth index than what we have at present. That is the reason why we study a survey like Euclid, that will allow us to obtain information not only about the matter distribution (power spectrum $P(k)$ of perturbations) but also about the weak lensing (WL) spectrum.

We perform a Fisher matrix analysis of the sensitivity of a Euclid-like survey to the growth index using the marginalized and 
complete power spectrum $P(k)$ and WL observables, and find that one can improve significantly its determination 
if we know the specific form of the redshift dependence of the growth index. We find that using $P(k)$ alone, 
one can reach (2\%, 5\%) errors in $(w_0, w_a)$, and (4\%, 12\%) errors in $(\gamma_0, \gamma_a)$, 
while using WL we get errors at least twice as large. These estimates allow us to differentiate easily between 
DGP, $f(R)$ models and $\Lambda$CDM, while it would be more difficult to distinguish the latter from a 
variable equation of state parameter or LTB models using only the growth index.

\section{The background equations}\label{s:background}

Here we review the basic equations for the relevant background quantities.
The evolution of the dark energy can be expressed by the present dark energy 
density $\Omega_{DE}$ and its equation of state parameter: 
\be
w(a) = \frac{p}{\rho}\,.
\ee  
Given any $w(a)$, the dark energy density is given by:
\be
\rho(a)=\rho_0a^{-3\left(1+\hat{w}\right)}
\ee
where 
\be
\hat{w}(a)=\frac{1}{\ln a}\int_{1}^{a}\frac{w\left(a'\right)}{a'}{\rm d}a'\,.
\ee
The Hubble parameter, $H(a)$, is
\be
H^2(a) = H_0^2\Big[\OM\,a^{-3} + \left(1-\OM\right)\,a^{-3\left(1+\hat{w}\right)}\Big]\,,
\label{eq:hubble}
\ee
where the subscript $0$ denotes the present epoch, and we are assuming global spatial flatness ($K=0$).
The angular diameter distance becomes
\bea \label{eq:angulardistance}
D_{A}(a) &\!=\!& a\int_{a}^{1}\frac{{\rm d}a'}{a'^2H(a')} 
= \frac{1}{1+z}\int_0^z\frac{dz'}{H(z')} 
= \frac{2}{H_0\sqrt{\OM}(1+z)}\times\\[1mm]
&\!\times\!& \left({}_2\!F_1\Big[\frac{1}{2},\frac{-1}{6\hat{w}},1-\frac{1}{6\hat{w}},1-\OM^{-1}\Big]-
\frac{1}{\sqrt{1+z}}{}_2\!F_1\Big[\frac{1}{2},\frac{-1}{6\hat{w}},1-\frac{1}{6\hat{w}},1-\Omega_m^{-1}(a)\Big]\right)\,.
\nonumber
\eea
The Eqs.~(\ref{eq:hubble}) and (\ref{eq:angulardistance}) imply that we limit ourselves to a flat universe which is filled with matter (denoted by the subscript $m$) and a general dark enery fluid.

The total matter and dark energy densities are:
\bea
\Om_{\rm m}(a) & = & \left(1 + \frac{1-\OM}{\OM}a^{-3\hat{w}}\right)^{-1} \,, \label{OM}\\
\Om_{\rm de}(a) & = & \left(1 + \frac{\OM}{1-\OM}a^{3\hat{w}}\right)^{-1} = 1 - \Om_{\rm m}(a)\,.
\label{OME}
\eea

\section{The perturbation equations}\label{s:pert}

In this section, we consider linear perturbations about a spatially-flat 
background model, defined by the line element:
\be
{\rm d}s^2 = a^2(\tau)\left[-\left(1+2\Psi\right){\rm d}\tau^2+\left(1-2\Phi\right){\rm d}x_{i}{\rm d}x^{i}\right]
\label{eq:newt-metric}
\ee
where $\tau$ is the conformal time.
From Eq.~(\ref{eq:newt-metric}) it is clear that we are working in 
the Conformal Newtonian (or longitudinal) gauge, which influences 
the evolution of the perturbations on big scales (especially those 
larger than the Hubble horizon). However, on smaller scales the 
choice of the gauge is less important, and we expect the 
perturbations to evolve independently from the gauge choice. 

The perturbation equations for a general fluid with an equation 
of state parameter $w =p/\rho$ are \cite{mabe,durrer}:
\bea
\delta' &=&3\left(1+w\right)\Phi'-\frac{V}{Ha^2}-3\frac{1}{a}\left(\frac{\delta p}{\rho}-w\delta\right)\,,
\label{eq:delta}\\
V' &=& -\left(1-3w\right)\frac{V}{a}+\frac{k^2}{Ha^2}\frac{\delta p}{\rho}+\left(1+w\right)\frac{k^2}{Ha^2}\Psi-\left(1+w\right)\frac{k^2}{Ha^2}\sigma\,,
\label{eq:V}
\eea
where $\delta =\delta\rho/\rho$ is the density contrast, $V=ik_{j}T^{j}_{0}/\rho$ is the 
scalar velocity perturbation (see \cite{ks1}) and the primes denote 
derivatives with respect to the scale factor $a$; also $\delta p$ and $\sigma$ are 
the pressure perturbation and the anisotropic stress, respectively, and 
they depend on the characteristics of the fluid itself.

In this paper we are interested on the evolution of the matter density field, 
which implies that $w=\delta p=\sigma =0$, at small scales (but sufficiently large 
for linear perturbation theory to still hold).
Furthermore, we also assume that the dark energy fluid enters only 
at the background level. This assumption is usually violated as 
dark energy perturbations may influence the evolution of the matter density field 
especially when the dark energy sound speed is small~\cite{aks,sk, ska}, since
it is then able to cluster at small scales. 
However, under the assumption of a high value of the 
dark energy sound speed (as is the case for the quintessence model, where 
$c_{s}^2=1$) we expect this assumption still to hold because 
dark energy is able to cluster only on very large scales.

Under these assumptions, Eqs.~(\ref{eq:delta}) and (\ref{eq:V}) 
read:
\bea
\delta'_{\rm m} &=&-\frac{V_{\rm m}}{Ha^2}\,,\label{eq:deltam}\\
V'_{\rm m} &=& -\frac{V_{\rm m}}{a}+\frac{k^2}{Ha^2}\Phi\label{eq:Vm}.
\eea
The above equations are linked together through the 
gravitational potential $\Phi$
\be
k^2\Phi=-4\pi Ga^2\rho_{\rm m}\left(\delta_{\rm m}+\frac{3aH}{k^2}V_{\rm m}\right)
\label{eq:potential-grav}
\ee
and $\Psi=\Phi$.

Joining Eq.~(\ref{eq:deltam}) and Eq.~(\ref{eq:Vm}), the master equation 
for the matter denstiy contrast becomes:
\be
a^2\delta''_m(a) +\big(3-\epsilon(a)\big)a\,\delta'_m(a) - \frac{3}{2}\Omega_m(a) \delta_m(a) = 0\,,
\label{eq:mastereq}
\ee
where $\epsilon(a) = -d\log H(a)/d\log a$. 
The exact growing mode solution of the above differential equation, for a 
constant dark energy equation of state parameter, $w$, is~\cite{lee, enqvist1}
\be
\delta_m(a) = a\cdot {}_2\!F_1\left[\frac{w-1}{2w},\frac{-1}{3w},1-\frac{5}{6w}; 
1-\Omega_m^{-1}(a)\right] \,,
\label{eq:solution}
\ee
where ${}_2\!F_1$ are Hypergeometric functions, see Ref.\cite{AS}, and
the full solution, including decaying modes, is reported in the Appendix \ref{app:fullsolution}.
Equation~(\ref{eq:solution}) can be further simplified using trasformation
formulae for the Hypergeometric functions~\cite{AS} into 
\be
\delta_m(a)=a\,\Omega_{m}(a)^{-\frac{1}{3w}}\ {}_2\!F_1\left[-\frac{1}{3w},
\frac{1}{2}-\frac{1}{3w},1-\frac{5}{6w}; 1-\Omega_{m}(a)\right]\,.
\label{eq:solution1}
\ee
It is worth having a further look at the structure of the solution found. Eq.~(\ref{eq:solution1}) 
is composed of two terms: the first one 
is the usual scale factor $a$, which is the solution we expect if dark energy had been 
neglected also at background level; furthermore, Eq.~(\ref{eq:solution1}) reduces to the 
classical solution in the matter domination era; this can be clearly seen if 
we set $\Omega_{m}(a)\rightarrow 1$, then the Hypergeometric function is $1$.
The second term instead, contains all the information about the 
dark energy fluid: via a direct dependence on $\Omega_{m}(a)$ and 
the Hypergeometric function, whose dependence is not straightforward. 
It is worth noticing that the slowed contribution to the dark matter comes from the 
term $\Omega_m(a)^{-\frac{1}{3w}}$ but its contribution is too big 
and it suppresses too much the growth of matter perturbations, 
the Hypergeometric function in Eq.~(\ref{eq:solution1}), instead, is a function that slowly increases 
with the scale factor, from $1$ for $a\ll1$ to just $1.16$ for $a\sim1$. This compensates for the 
extra suppression of the $\Omega_m(a)^{-\frac{1}{3w}}$ term.

In order to evaluate the growth rate, we start using Eq.~(\ref{eq:solution1}), and write it as, see Appendix A, 
\be\label{faexact}
f(a)\, = \,\frac{d\log\delta_{m}}{d\log a}
\, = \,\Om^{1/2}_{m}(a)\frac{P_{1/6w}^{5/6w}\left[\Om^{-1/2}_m(a)\right]}
{P_{-1/6w}^{5/6w}\left[\Om^{-1/2}_m(a)\right]}\,.
\ee
This function is simpler to evaluate for a given model, and has the explicit dependence of the growth 
factor as a function of $\Om_m(a)$. We will use it in the next subsection on the gamma parametrization.

\subsection{Varying equation of state parameter}

We can extend our discussion also to a varying dark energy equation of state parameter $w(a)$. In this 
case there are no exact analytical solutions for the matter density contrast as for the case of a constant $w$;
the main problem is that there is no direct transformation between the scale factor $a$ and the new 
variable $u$, see Appendix A. Here we assume the equation of state parameter to be:
\be
w(a) = w_0+w_a\,(1-a)\,,
\ee
for which the matter density parameter can be integrated
\be
\Om_m(a) = \left(1 + \frac{\ODE}{\OM}a^{-3(w_0+w_a)}\,e^{3w_a(a-1)}\right)^{-1} \,.
\ee
We can also integrate the master equation in this case, and although we have found no exact analytical solution, 
we found an approximate solution which is within 0.1\% of the numerical solution for the whole range of $(w_0,w_a)$ values.
In this case the growth rate reads
\be
f(a)\ =\ \Om^{1/2}_{m}(a)\ \frac{P_{1/6w(a)}^{5/6w(a) + w_a a/6w^2(a)}
\left[\Om^{-1/2}_{m}(a)\right]}{P_{-1/6w(a)}^{5/6w(a) + w_a a/6w^2(a)}
\left[\Om^{-1/2}_{m}(a)\right]}\,.
\label{fwa}
\ee
This is a very compact expression which will be very useful in the next subsection.

\subsection{The gamma parametrization}\label{s:gamma}

A potentially more promising place to look for the impact of dark energy 
is the way in which the matter perturbations grow over time. 
Often the impact of the dark energy on the growth rate of the matter perturbations is 
parametrized in terms of the growth index $\gamma$, defined through \cite{peebles}:
\be
f(a)\, =\, \Omega_m(a)^{\gamma}\,.
\label{eq:growth-rate}
\ee
Such a parametrization assumed $\gamma$ to be constant and, moreover, with a value that was very approximately 
$\gamma\simeq0.6$ for general relativity plus a cosmological constant $\Lambda$, i.e. for a constant equation 
of state parameter $w=p/\rho=-1$. Furthermore, detailed studies went further and computed the $\gamma$ 
parameter for a constant but arbitrary $w$, \cite{lahav}:
\be
\gamma = \frac{3(w-1)}{(6w-5)}\,,
\label{eq:gamma-const}
\ee
which reduces to $\gamma=6/11\simeq0.55$ for $w=-1$. The last equation does not take into 
account the dependence of $\gamma$ on the matter density parameter. There is an expression for the 
growth index $\gamma$ which does depend on $\Omega_m(a)<1$, to first order, see Ref.~\cite{wast},
\be
\gamma = \frac{3(w-1)}{(6w-5)}+\frac{3}{2}\frac{\left(1-w\right)\left(2-3w\right)}
{\left(5-6w\right)^3}\big(1-\Om_m(a)\big)+\OO\big(1-\Om_m\big)^2\,,
\label{eq:gamma-wast}
\ee
which reduces to the well known result (\ref{eq:gamma-const}) in the limit $a\to0$. However, we have found in Eq.~(\ref{faexact}) an exact solution {\em for all $a$}. Using Eq.~(\ref{eq:growth-rate}) we have
\be
\gamma(a) = \frac{1}{2}+\frac{1}{\ln\Om_{m}(a)}\ln\left[\frac{P_{1/6w}^{5/6w}\left[\Om^{-1/2}_m(a)\right]}{P_{-1/6w}^{5/6w}\left[\Om^{-1/2}_{m}(a)\right]}\right]\,.
\label{eq:gamma-2f1}
\ee
In particular, Eq.~(\ref{eq:gamma-2f1}) depends not only on $w$, but also on $\OM$. 
For the fiducial values $w=-1$ and $\OM=0.25$, one finds $\gamma(a=1)=0.556$ instead of $\gamma(a\rightarrow 0)=0.545$. 
For present day purposes, with galaxy surveys providing at most a few percent accuracy on the growth parameter, this difference $-$ of order 3\% $-$ may seem academic. However, for future surveys like PAU \cite{pau}, LSST \cite{lsst} or Euclid~\cite{euclid}, where we will have tomographic reconstruction of the past history in both the matter distribution and the expansion rate, up to redshift $z\simeq2$, these differences may begin to play an important role as a discriminator between standard GR with a cosmological constant and e.g. modified gravity theories like $f(R)$, or quintessence models.

The same discussions apply when we want to evaluate the growth index when the equation of state parameter 
is a function of the scale factor,
\be
\gamma(a) = \frac{1}{2}+\frac{1}{\ln\Om_{m}(a)}\ln\left[\frac{P_{1/6w(a)}^{5/6w(a)+w_aa/6w^2(a)}\left[\Om^{-1/2}_{m}(a)\right]}{P_{-1/6w(a)}^{5/6w(a)+w_aa/6w^2(a)}\left[\Om^{-1/2}_m(a)\right]}\right]\,.
\label{eq:gamma-2f1-wa}
\ee
In the next section we will discuss the growth index for different Dark Energy models.

\section{The growth index for different Dark Energy models}

Here we discuss a range of models of dark energy which have very different behaviours for the growth index $\gamma$ as a function of redshift. The models we consider here are $f(R)$ theories, the DGP model, LTB cosmologies and $wCDM$. With this choice of models we wish to cover the different possibilities to explain dark energy: modified gravity theories ($f(R)$ and DGP models), modified geometry (LTB models) and modified matter through $w$CDM, which can be associated with, for example, a quintessence model through the correspondence between $f(\phi)$ and $w(a)$.

\subsection{The Dvali-Gabadadze-Porrati model}

The brane-world scenario offers an alternative to explain the current acceleration of the universe. In the DGP model \cite{dgp}, gravity lives in a 4 dimensional brane with the rest of matter fields for scales $\lambda<r_c$ while it seeps into a fifth dimensional bulk for $\lambda>r_c$, weakening its strength in our brane. For such models, the Hubble parameter is given by \cite{maartens}
\be
H(a)=H_0\left[\sqrt{\Omega_{r_c}} +\sqrt{\Omega_{r_c} +\OM \ a^{-3}}\right]
\ee
where $\Omega_{r_c}=1/(4r_c^2H_0^2)=(1-\OM)^2/4$. The Poisson equation is modified in these models as follows:
\be\label{eq:phidgp}
k^2\Phi = -\frac{\kappa^2}{2}\left(1-\frac{1}{3\beta}\right)\rho_m\delta_m
\ee
where $\beta=1-\frac{2(Hr_c)^2}{2Hr_c-1}$. For the growth index, we have used the approximation found by \cite{linder, nesseris}:
\be
\gamma(a) =\frac{7 + 5\Omega_{\text{M}}(a) + 7\Omega^2_{\text{M}}(a) + 3\Omega^3_{\text{M}}(a)}{
\left[1 + \Omega^2_{\text{M}}(a)\right]\left[11 + 5\Omega_{\text{M}}(a)\right]}\,,
\label{eq:dgpapprox}
\ee
where $\Omega_{\rm M}$ is given by
\be
\Om_m(a) = 1-\frac{1}{Hr_c}=\frac{\left[1+\frac{4\OM}{a^3(1-\OM)^2}\right]^{1/2}-1}{\left[1+\frac{4\OM}{a^3(1-\OM)^2}\right]^{1/2}+1}\,,
\ee
while the effective equation of state reads
\be
w(a) = \frac{Hr_c}{1-2Hr_c} = \frac{-1}{1+\Om_m(a)}\,.
\ee

\subsection{The Starobinsky $f(R)$ model}

The simplest way to modify general relativity is by substituting the Ricci scalar $R$ with a function $f(R)$, in the Einstein-Hilbert action, \cite{starobinsky1980}:
\be
S=\frac{1}{2\kappa^2}\int d^4x\sqrt{-g}f(R)+S_m(g_{\mu\nu},\Psi_m)
\ee
The equivalent of the Friedmann equations for a perfect nonrelativistic matter fluid for this action are given by
\be
\begin{aligned}
3FH^2&=\rho_m + \frac{FR-f(R)}{2}-3H\dot F,\\
-2F\dot H&=\rho_m+\ddot F -H\dot F,
\end{aligned}
\ee
where $F\equiv f'(R)$. The Ricci scalar $R$ is expressed in terms of the Hubble parameter as $R=12H^2+6\dot H$ for a flat background. It is convenient to introduce a set of dimensionless variables to study the dynamics of $f(R)$ gravity~\cite{agpt}:
\be
x_1=-\frac{\dot F}{HF}, \qquad x_2=-\frac{f(R)}{6FH^2}, \qquad x_3=\frac{R}{6H^2}, \qquad x_4=aH.
\ee
The matter density parameter is then given by
\be
\tilde{\Omega}_m\equiv\frac{\rho_m}{3FH^2}=1-x_1-x_2-x_3
\ee
With these variables we obtain the following closed system of differential equations
\be
\begin{aligned}
x_1'&=-1-x_3-3x_2+x_1^2-x_1x_3\,,\\[2mm]
x_2'&=\frac{x_1x_3}{m(r)}-x_2(2x_3-4-x_1)\,,\label{eq:coupled_fR}\\
x_3'&=-\frac{x_1x_3}{m(r)}-2x_3(x_3-2)\,,\\
x_4'&=(x_3-1)x_4\,.
\end{aligned}
\ee
where a prime represents a derivative with respect to $N=\ln a$ and
\be
m(r)\equiv\frac{Rf_{,RR}}{f_{,R}}, \qquad r\equiv-\frac{Rf_{,R}}{f}=\frac{x_3}{x_2}.
\ee
The linear perturbation equation for the matter density contrast in the comoving gauge (where the matter velocity vanishes), in terms of the dimensionless variables, $x_1,\ x_2$ and $x_3$, is \cite{tsujikawafr}
\be
\begin{aligned}
\delta''_m+(x_1+x_3)\delta'_m-3(1-x_1-x_2-x_3)\delta_m= \left[3\left(x_1+x_3-\frac{x_3}{m}-1\right)-\frac{k^2}{x_4^2}\right]\delta\tilde F -3\delta\tilde F'\,,
\label{eq:difeq_fR}
\end{aligned}
\ee
where the differential equation for $\delta \tilde{F}=\delta F/F$ is
\be
\delta\tilde{F}''+(2x_1-x_3-1)\delta\tilde{F}'+\left[\frac{k^2}{x_4^2}-x_3+\frac{2x_3}{m}+3x_2-x_1+1\right]\delta\tilde{F}=0\,.
\label{eq:difeq_dF}
\ee
Solving the coupled differential equations (\ref{eq:coupled_fR}), we can obtain the background functions 
\be\begin{aligned}
H(a) &= \frac{x_4(a)}{a}\,,\\[2mm]
\Omega_m(a) &= F(a)\big(1-x_1(a)-x_2(a)-x_3(a)\big)\,,\\[1mm]
w(a) &= \frac{1 - 2\, x_3(a)}{3(1 - \Omega_m(a))}\,.
\end{aligned}
\ee
Then solving equations (\ref{eq:difeq_fR}) and (\ref{eq:difeq_dF}) numerically for a given scale $k$, we find a solution for $\delta_m$, and compute from it the growth index $\gamma$.

In this paper we study the Starobinsky model~\cite{starobinsky} where $f(R)$ is given by
\be
f(R)=R+\lambda R_0\left[\left(1+\frac{R^2}{R_0^2}\right)^{-n}-1\right]
\ee
where $\lambda$ and $n$ are two positive constants and $R_0$ corresponds to the present value of the Ricci scalar. To be in agreement with observations, we take $n=2$ and $\lambda=2$ \cite{starobinskyobs}. We also take $k=0.16\,h/$Mpc in Eq.~\eqref{eq:difeq_fR} to calculate the density contrast and the growth index.

\subsection{The Lema\^{i}tre-Tolman-Bondi large-void model}

A plausible alternative to explain the current acceleration of the universe are inhomogeneous universe models in which the acceleration we seem to perceive is caused by our position as observers inside an underdense region of space. One of the simplest models to study the effect of such large inhomogeneities is the spherically symmetric Lema\^{i}tre-Tolman-Bondi model \cite{lemaitre, tolman, bondi}. In this model, the metric is given by
\be
ds^2 = - dt^2 + X^2(r,t)\,dr^2 + A^2(r,t)\,d\Omega^2\,,
\ee
where $d\Omega^2 = d\theta^2 + \sin^2\theta d\phi^2$ and the equivalent of the scale factor now depends on the radial coordinate as well as time. We can find a relationship between $X(r,t)$ and $A(r,t)$ using the the $0-r$ component of the Einstein equations: $X(r,t)=A'(r,t)/\sqrt{1-k(r)}$ where a prime denotes a derivative with respect to coordinate $r$ and $k(r)$ is an arbitrary function that plays the role of the spatial curvature parameter.

To find the growth index in LTB cosmologies we must study linear perturbation theory in inhomogeneous universes. Due to the loss of a degree of symmetry, the decomposition theorem does no longer hold. This means that, in general, our perturbations will no longer decouple into scalar, vector and tensor modes. A study of the perturbation equations in this scenario using a 1+1+2 decomposition of spacetime can be found in \cite{clarkson}. However, if the normalized shear $\varepsilon=(H_T-H_L)/(2H_T+H_L)$ is small, as observations seem to confirm \cite{shear}, we can use the ADM formalism and express our perturbed LTB metric as
\be
ds^2 = -(1 + 2\Phi)dt^2+ (1-2\Psi)\gamma_{ij}dx^idx^j
\ee
where $\gamma_{ij}=\text{diag}\{X^2(r,t),A^2(r,t),A^2(r,t)\sin^2\theta\}$. Within this formalism, the evolution equation for a pressure-less fluid in the conformal Newtonian gauge (where the absence of anisotropic stresses gives $\Phi=\Psi$) is given by
\be
\ddot\Phi+4H_T\dot\Phi+(4\dot H_T+6H_T^2)\Phi=0
\label{eqphi}
\ee
where we now have two different expansion rates $H_T(r,t)=\dot A/A$ and $H_L(r,t)=\dot A'/A'$ which correspond respectively to the transverse and longitudinal expansion rates. The growing mode solution of equation \eqref{eqphi} is given by 
\be
\Phi(r, t) = \Phi_0(r)\ \!{}_2\!F_1\Big[1, 2, \frac{7}{2}; u\Big]\,,
\ee
where $u = k(r)A(r, t)/F(r)$ and $F (r) = H_0^2(r)\Omega_M(r)A^3(r,t_0)$ specifies the local matter density today. With this solution we find the density contrast, 
\be
\delta(r,t)=\frac{A(r,t)}{r}\Phi(r,t)
\ee
We can now calculate the growth index as before, noting that now the matter density parameter is a function of redshift via both time $t$ and the radial coordinate $r$. In LTB models, this is in principle an arbitrary function which must be fixed in each case. In the case of the constrained GBH model \cite{GBH,kSZ} the parameters are given by
\begin{eqnarray}
\Omega_M(r) &\!=\!& 1 + (\Omega_M^{(0)}-1){1-\tanh[(r-r_0)/2\Delta r]\over1+\tanh[r_0/\Delta r]}\\
H_0(r) &\!=\!& H_0\left[{1\over1-\Omega_M(r)} - {\Omega_M(r)\over(1-\Omega_M(r))^{3/2}}
{\rm arcsinh}\sqrt{1-\Omega_M(r)\over\Omega_M(r)}\right]\,,
\end{eqnarray}
with
\begin{equation}
r_0 = 3.0\ {\rm Gpc}\,, \hspace{5mm} \Delta r = 1.5\,r_0\,,  \hspace{5mm}
h_0 = 0.74\,,  \hspace{5mm} \Omega_M^{(0)} = 0.15\,,
\end{equation}
where these values have been chosen to best fit the supernovae data \cite{shear,grande}.
Within this model, the growth rate, i.e. the logarithmic derivative of the density contrast, is given by
\be\label{fzLTB}
f(z)=1 + \frac{4}{7} \Big(1 - \Omega^{-1}_m(z)\Big) 
\frac{{}_2\!F_1\Big[2, 3, \frac{9}{2}; 1 - \Omega^{-1}_m(z)\Big]}{{}_2\!F_1\Big[1, 2, \frac{7}{2}; 1 - \Omega^{-1}_m(z)\Big]}\,,
\ee
where $\Omega_m(z)$ is the fraction of matter density to critical density, as a function of redshift.\footnote{The matter density in LTB model is given by $\rho(r,t)=F'(r)/A'(r,t)A^2(r,t)$. Note that this is different from $\Omega_M(r)=F(r)/A^3(r,t_0)H_0^2(r)$, which gives the mass radial function today, see Ref.~\cite{GBH}.} This function (\ref{fzLTB}) is identical to the {\em instantaneous} growth function of matter density in an open universe, where the local matter density $\Om_M$ is given by $\Omega_m(z)$ at that redshift. This is a good approximation only in LTB models with small cosmic shear,  see Ref.~\cite{AGBHV}.

Alternatively, we can write the growth function in terms of Legendre polynomials,
\be
f(z)=\Omega^{1/2}_m(z) \frac{P_{-1/2}^{-5/2}\Big[\Omega^{-1/2}_m(z)\Big]}{P_{1/2}^{-5/2}\Big[\Omega^{-1/2}_m(z)\Big]}\,,
\ee
which can also be written in terms of ordinary functions,
\be
f(z) = \frac{9u(1-u^2)+6(1+2u^2)\sqrt{u^2-1}\ {\rm arcsinh}\sqrt{(u-1)/2}}
{2u(u^4+u^2-1)-12u^2\sqrt{u^2-1}\ {\rm arcsinh}\sqrt{(u-1)/2}} \,,
\ee
where $u\equiv\Omega^{-1/2}_m(z)>1$.

The equation of state $w(z)$ has been obtained for this model from the expression
\be
w(a) = \frac{d\log(\Omega^{-1}_m(a)-1)^{-1}}{d\log a^3} = \frac{a\,\Omega_m'(a)/\Omega_m(a)}{3(1-\Omega_m(a))}
\label{wLTB}
\ee
which was used in Fig.~\ref{fig:compared_DEmodels}. Note that the rate of expansion $H(z)$ for this model is similar to that of $\Lambda$CDM, which explains why it fits the SNIa data \cite{enqvist, celerier}.

\subsection{Comparison of the growth index for the different models}

\begin{figure}[h!]
  \centering
  \subfloat[Hubble rate]{\label{fig:Hubble_all}\includegraphics[width=0.5\textwidth]{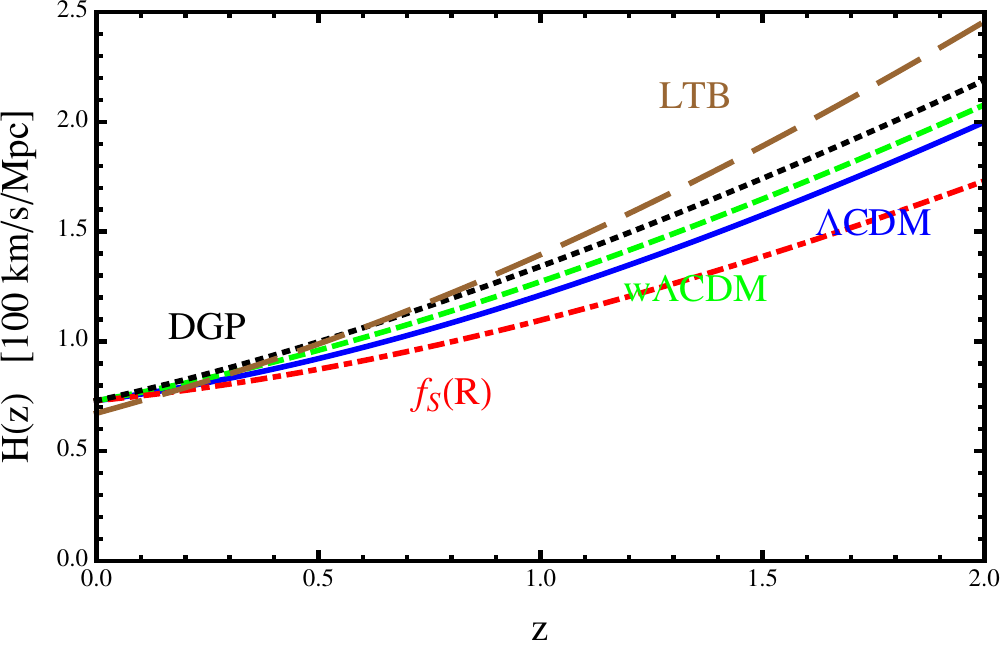}}
  \subfloat[Matter density]{\label{fig:omegam_all}\includegraphics[width=0.5\textwidth]{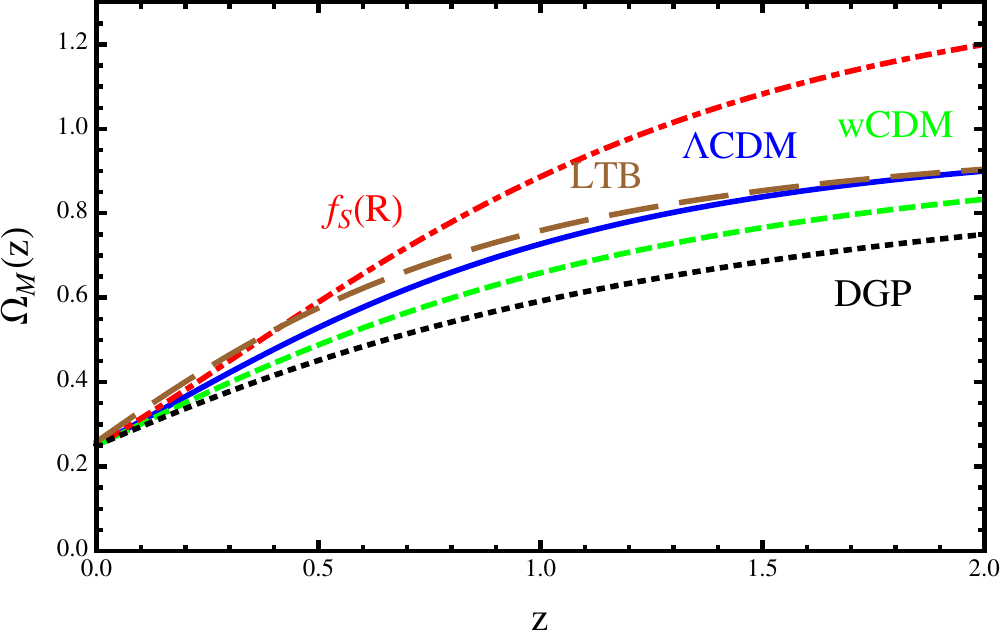}}\\
  \hspace{-0.2cm}\subfloat[Equation of state]{\label{fig:wa_all}\includegraphics[width=0.51\textwidth]{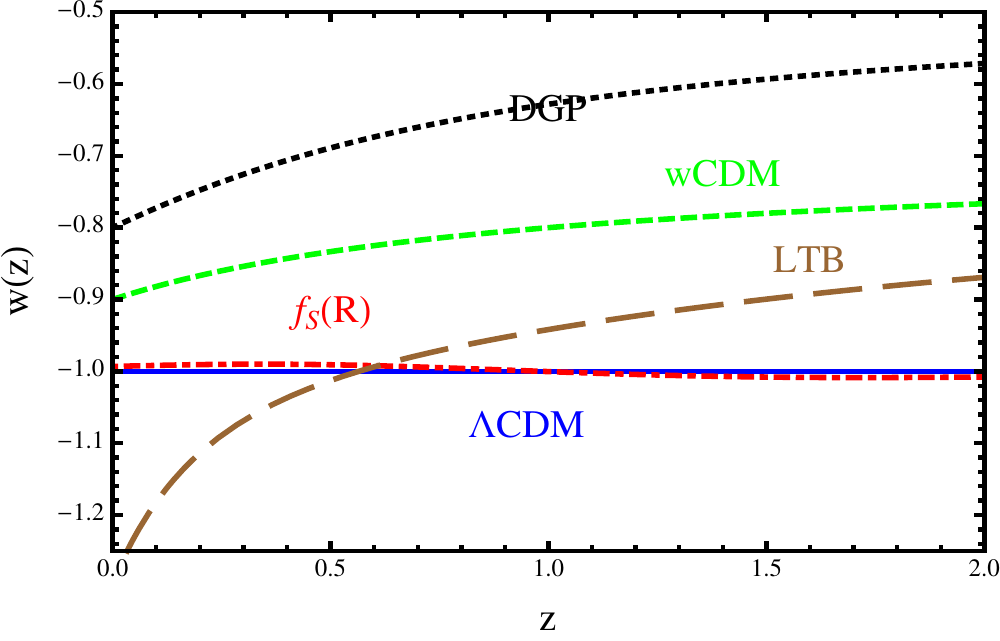}}
  \subfloat[Growth index]{\label{fig:gamma_all}\includegraphics[width=0.5\textwidth]{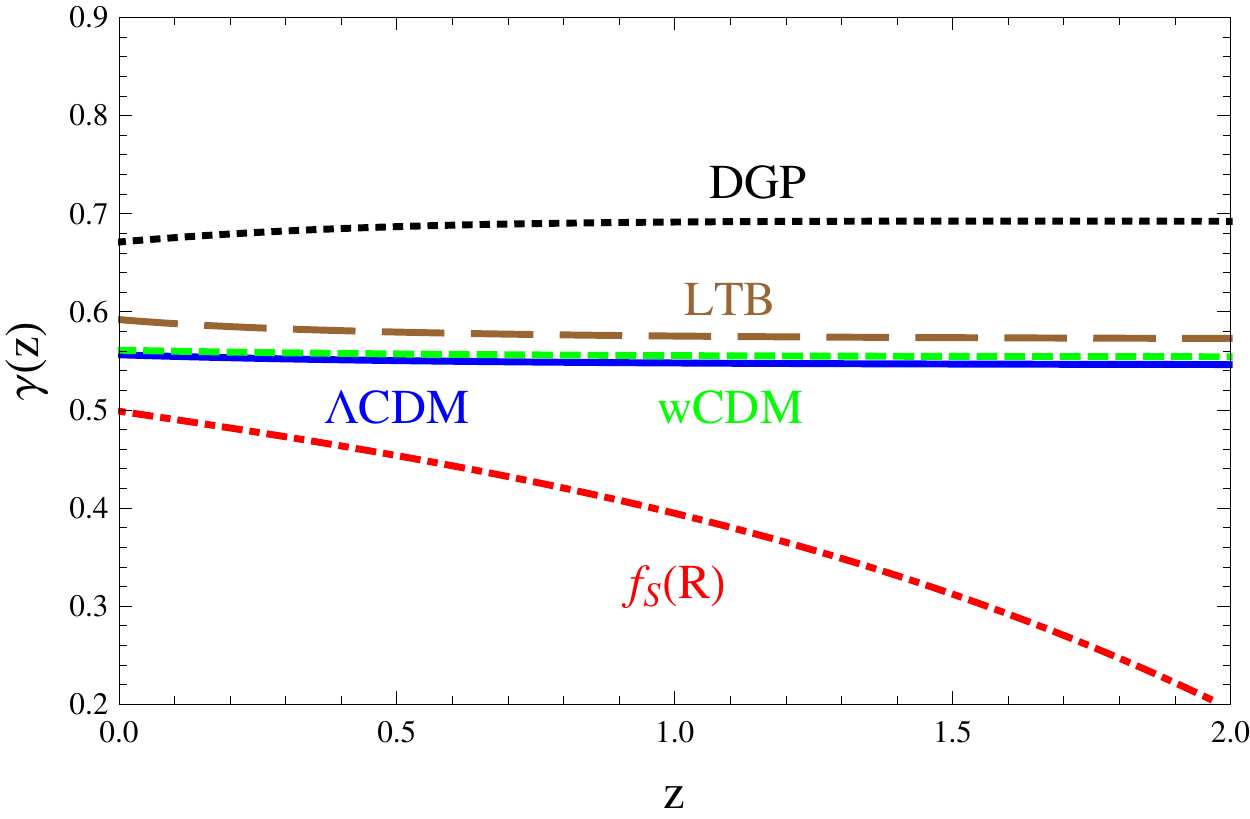}}
  \caption{Hubble rate, matter density parameter, equation of state and growth index as a function of redshift for different DE models. The solid lines represent the $\Lambda$CDM model, the short dashed lines the wCDM one, the dotted lines the DGP model, the dotted dashed lines the $f(R)$ theory and the long dashed lines, the LTB model.}
  \label{fig:compared_DEmodels}
\end{figure}

Now we are ready to compare the results of the different models studied here. In Fig.~\ref{fig:compared_DEmodels} we can see the Hubble rate, the matter density parameter, the equation of state and the growth index for all the models studied. In the case of wCDM we have taken $w_0=-0.9$ and $w_a=0.2$. As it is seen, even though the expansion history, matter density parameter and equation of state are quite different for the different models considered, most of them have a similar growth index, being the biggest difference between DGP, $f(R)$ and the rest.

This difference in growth indices comes from how linear perturbations in the matter density are described in each of the models. In the case of $f(R)$, if we neglect the oscillation mode of $\delta F$ relative to the mode induced by matter perturbations $\delta_m$, we can obtain the following approximate equation for matter perturbations:
\be
\ddot \delta_m+2H\dot\delta_m-4\pi G_{\rm eff}\rho_m\delta_m\simeq0
\ee
where
\be
G_{\rm eff}\equiv\frac{G}{F}\left(\frac{1+4\frac{k^2F'}{a^2F}}{1+3\frac{k^2F'}{a^2F}}\right)
\label{eq:geff}
\ee
where here a prime means a derivative with respect to $R$, and $G$ is Newton's gravitational constant.
The quantity $G_{\rm eff}$ encodes the modification of gravity due to the presence of the scalaron field. Thus, the Poisson equation in Fourier space is transformed by replacing Newton's gravitational by the effective one in Eq.~\eqref{eq:geff}. The transition from the GR regime to the scalar-tensor one occurs when $m\sim(aH/k)^2$. For the wave numbers relevant to the observable linear region of the matter power spectrum we require $m(z=0)\gtrsim3\times 10^{-6}$ for the transition to have occurred by today. The Starobinsky model in particular allows for a rapid growth of $m$ from $R\gg H_0^2$ ($m\lesssim 10^{-15}$) to $R\simeq H_0^2$ ($m=\mathcal{O}(0.1)$).

Another way to understand the evolution of $\gamma$ in the Starobinsky model is analysing the evolution equation \cite{polarski}. At high redshifts, we can approximate this evolution equation to \cite{MSY}
\be
\begin{aligned}
&(1+z)(1-\Omega_m)\frac{d\gamma}{dz}\\
&=\frac{3}{2}\left(\frac{G_{\rm eff}}{G}-1\right)+(1-\Omega_m)\left[\frac{11}{2}\left(\gamma-\frac{6}{11}\right)-\frac{3}{2}(1-\gamma)\left(\frac{G_{\rm eff}}{G}-1\right)-\frac{3}{2}(2\gamma-1)(w_{\text{DE}}+1)\right]
\end{aligned}
\label{evogamma}
\ee
At early stages, the first term of the right hand side of Eq.~\eqref{evogamma} dominates. This is why $\gamma$ decreases as $G_{\rm eff}/G$  increases, even becoming negative. As we approach the present era, the second term starts to dominate making $\gamma$ increase again.

In the case of the DGP model, the effect of the extra dimension affects both the friction term in the evolution equation for $\delta_m$ and the source term as, similarly to $f(R)$ gravity, we can define an effective gravitational constant as
\be
G_{\rm eff}=G\left(1-\frac{1}{3\beta}\right)
\ee
where here $\beta$ is the same as the parameter described in Eq.~\eqref{eq:phidgp}. As we can see in Fig.~\eqref{fig:omegam_all}, the parameter $\Omega_{\rm M}$ is smaller for the DGP model than for the standard $\Lambda$CDM one. To compensate for this lack of matter density, the growth index is higher than for the rest of the models, as it is clear from Fig.~\eqref{fig:gamma_all}.

In the case of the LTB model, we see in Fig.~\eqref{fig:wa_all} that the equation of state parameter $w$ differs significantly from all the other models. The reason for this can be seen by looking at the formula used to calculate $w(a)$, Eq.~\eqref{wLTB}. The expansion rate is similar to the $\Lambda$CDM case since the parameters chosen for the void model considered have been chosen to fit the supernovae data. As we can see, the growth index for this model is not significantly different from the flat universe scenario and also tends to an asymptotic value given by $\gamma=0.573$, very close to the predicted one in the $\Lambda$CDM case.

As regards the wCDM model studied, we have used the values of $w_0=-0.9$ and $w_a=0.2$ to calculate the growth index using Eq.~\eqref{fwa}. We have chosen these values to see how a considerably different model from the standard $\Lambda$CDM one can have a growth index that is barely distinguishable from it, as it can be seen in Fig.~\eqref{fig:gamma_all}.

\subsubsection{Ansatz for the growth index}\label{subsubsec:ansatz}

\begin{figure}[t!]
\centering
\includegraphics[scale=1.0]{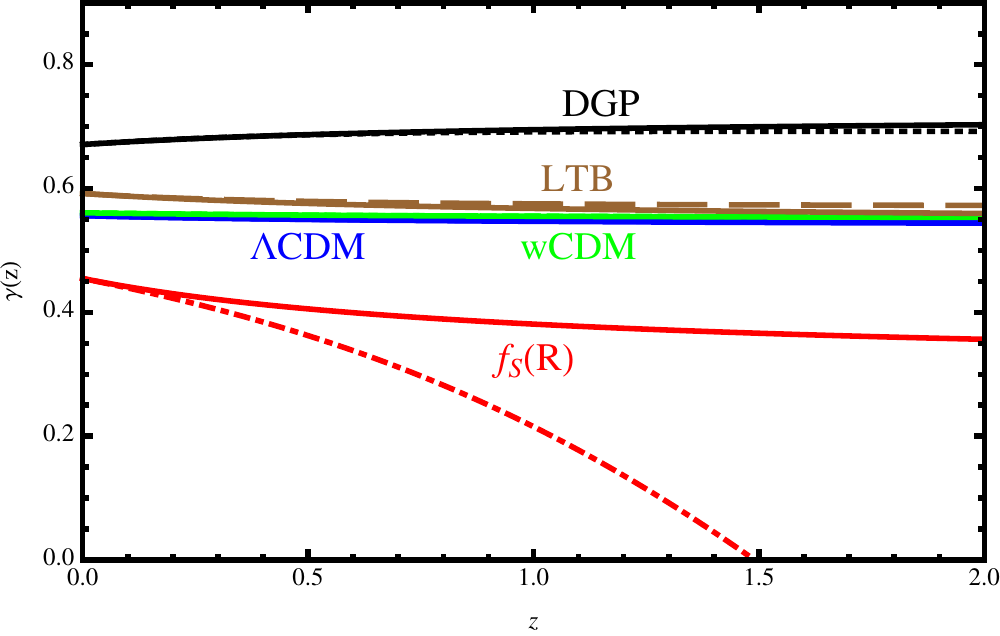}
\caption{The redshift evolution of the growth index compared to the linear approximation in Eq.~(\ref{gammaansatz}) (continuous lines), for all four classes of models. Note that while for most models the Ansatz gives an excellent description, for $f(R)$ models it deviates significantly even for moderate redshifts.}
\label{fig:gamma-ansatz}
\end{figure}

In this subsection we compare the different models' growth indices' redshift evolution with the Ansatz~\cite{portoamendola,wu}
\be\label{gammaansatz}
\gamma(a) = \gamma_{0}+\gamma_{a}(1-a) \,.
\ee
We have plotted in Fig.~\ref{fig:gamma-ansatz} the known redshift dependence of the growth index for the different models together with the corresponding values derived with the above Ansatz. As can be seen very clearly, for most models this is an excellent approximation. However, for $f(R)$ models it fails to work at relatively nearby redshifts. For this reason we have adapted our Fisher matrix analysis for this specific case, see Section~ \ref{sec:results}.

\section{Fisher matrix analysis prospects for a survey like Euclid}\label{sec:fisher}

The Fisher information matrix can be used to quantify the amount of information that an observable random 
variable $X$ carries about a set of unknown parameters $\theta$. 

More specifically, let us assume that we have a data set arranged as a n-dimensional vector 
$X=\{x_1,x_2,...x_n\}$ which can be thought of as a random variable with its 
probability distribution function $L\left(X; \theta\right)$ which depends on 
the model parameters $\theta$ in a known way. 
The Fisher matrix is defined as the second derivatives of the logarithm of 
the probability distribution function with respect the parameters:
\be
F_{ij}=\left\langle \frac{\partial^2\mathcal{L}}{\partial\theta_{i}\partial\theta_{j}}\right\rangle
\label{eq:fisher}
\ee
where $\mathcal{L}=-\ln L$. For a sufficiently large sample (in cosmology, a large survey), the 
central limit theorem ensures that the probability distribution function of the random 
variable $X$ becomes Gaussian centered about their true value. 
In this case, in Eq.~(\ref{eq:fisher}), the logarithm of the likelihood function $L\left(X;\theta\right)$ becomes 
a quadratic function of the variable $X$ about its maximum. So, only under the assumption 
of a large sample, the inverse of the Fisher matrix becomes the covariance matrix and 
it can be used to estimate the errors on the parameters and furthermore 
to look at degeneracies between them.  
Then, with the help of the Cramer-Rao inequality, 
one can estimate the minimum variance of the parameter from the data set: 
the root square of the diagonal elements of the inverse of the Fisher matrix give 
the best errors on the parameters, ie. $\Delta\theta_{i}=\big(F^{-1}\big)_{ii}$.

This method is extremely useful, for instance, when we want to optimize 
the design of a survey, in the sense that it is possible to model such a survey in order to 
maximize its information content on the particular parameters we want to measure.

\subsection{Complete galaxy power spectrum}

There are two ways we can get information about dark energy parameters by looking at the galaxy power spectrum: 
first, by marginalizing over the main observables: the Hubble parameter, the angular diameter distance, the growth factor 
and the redshift space distortion parameter and projecting on to the dark energy parameter space;
secondly, by taking the full information of the observed galaxy power spectrum and deduce from the latter the 
parameters of interest.

Following \cite{se,aqg,sa} we write the observed galaxy power spectrum as:
\begin{equation}
P_{\rm obs}\left(z;k,\mu\right) =  \frac{D_{Ar}^{2}(z)H(z)}{D_A^2(z)H_r(z)}G^2(z)b(z)^2
\left(1+\beta\mu^{2}\right)^2 P_{0r}(k) +P_{\rm shot}(z)
\end{equation}
where the subscript $r$ refers to the values assumed for the reference
cosmological model, i.e. the model at which we evaluate the Fisher
matrix, $b(z)$ is the matter bias factor, $G(z)$ is the growth factor of matter 
perturbations. Also here $P_{\rm shot}$ is the shot noise due to discreteness in
the survey, $\mu$ is the direction cosine within the survey, $P_{0r}$
is the present matter power spectrum for the fiducial (reference) cosmology. 
For the linear matter power spectrum we adopt the CAMB output \cite{camb}.

The wavenumber $k$ is also to be transformed between the fiducial
cosmology and the general one (see \cite{se, aqg, sa} for more details). 

The distortion induced by redshift can be expressed in terms 
of the $\beta(z)$ factor which is related to 
the bias factor via:
\begin{equation}
\beta(z)=\frac{\Omega_m(z)^{\gamma}}{b} = \frac{f(z)}{b}\,.
\label{eq:beta}
\end{equation}
The $\left(1+\mu^2\beta\right)^2$ factor accounts only for linear distortion in the 
redshift space and it should be considered as a first 
approximation. More work beyond Kaiser's small-angle and the Gaussian approximation
is needed, as discussed in \cite{hamilton, zaurobi, tegmark, scoccimarro}.

The  galaxy over-densities are assumed to trace the underlying matter distribution 
through a fraction called bias factor, $b(z,k)$. This quantity could be arbitrary, it could even
depend on both time and scale, see \cite{bias1, bias2, bias3}. Usually it is assumed that the bias on large scales 
is independent on scale \cite{bias4}, hence in the matter power spectrum, this term appears as a 
multiplicative factor which modulates the overall amplitude of the galaxy power spectrum. 
We will assume here a Gaussian linear bias, with redshift 
dependence $b(z)=\sqrt{1+z}$, because it provides a good fit to the $H_\alpha$ line galaxies in the 
near-infrared, which are the target of a Euclid-like survey, see \cite{biask}.

The factor $G\lzr$ is related to the growth rate by $f(a)=\frac{d\log G}{d\log a}-1$
and it is usually assumed to be independent on scale: 
the late-time change in the expansion rate affects all scales equally. 
However, the growth factor may depend on the scale $k$, for instance 
allowing perturbations also in the dark energy sector. In the last case, the 
existence of the dark energy sound horizon will introduce a $k$ dependence on the growth factor, 
see for instance \cite{sk, ska,growth-k}. 
However, here we assume a scale independent $G\lzr$ as we have neglected 
the dark energy perturbations.

The total galaxy power spectrum including the errors 
on the redshift can be written as \cite{se}
\begin{equation}
P(z,k) = P_{\rm obs}(z,k)e^{k^2 \mu^2 \sigma_r^2}
\end{equation}
where $\sigma_r=\delta z/H(z)$ is the absolute error 
on the measurement of the distance and $\delta z$ is the absolute error on redshift. 

The Fisher matrix provides a useful method for evaluating the marginalised errors on 
cosmological parameters.
Assuming the likelihood function to be Gaussian, the Fisher matrix
is \cite{ehut,tegmark97}
\begin{equation}
F_{ij}=2\pi\int_{k_{\rm min}}^{k_{\rm max}}\frac{\partial\log P\left(k\right)}{\partial\theta_{i}}
\frac{\partial\log P\left(k\right)}{\partial\theta_{j}}\cdot V_{\rm eff}\cdot\frac{k^{2}}{8\pi^{3}}\cdot {\rm d}k
\label{eq:FisherMatrix}
\end{equation}
where  the $\theta$'s are the parameters shown in Tab.~\ref{tab:cosmological-parameters};
the derivatives are evaluated at the parameter values of the
fiducial model and $V_{\rm eff}$ is the effective volume of the survey,
given by:
\begin{equation}
V_{\rm eff} =  \int {\rm d}^3\vec{r}\left(\frac{n\left(\vec{r}\right)P\left(k,\mu\right)}{n\left(\vec{r}\right)
P\left(k,\mu\right)+1}\right)^{2}= \left(\frac{\bar n\,
P\left(k,\mu\right)}{\bar n\,P\left(k,\mu\right)+1}\right)^{2}V_{\rm survey}.
\label{eq:Volume}
\end{equation}
Here $\mu=\vec{k}\cdot\widehat{r}/k$,~$\widehat{r}$ is the unit vector along the line of sight and $k$
the wave vector, and the last equality holds for an average comoving number density
$\bar n$. The highest frequency, $k_{\rm max}(z)$, is evaluated at 
$z$ of the corresponding bin and it is chosen 
so as to avoid non-linearity problems both in the spectrum
and in the bias; we choose values from $0.11\,h$/Mpc 
for small $z$ bins to $0.25\,h$/Mpc for the highest redshift
bins. 

Any submatrix of $F_{ij}^{-1}$ gives the correlation matrix
for the parameters corresponding to rows and columns on that submatrix. 
The eigenvectors and eigenvalues of this correlation matrix give the
orientation and the size of the semiaxes of the ellipsoid confidence region. 
This automatically marginalizes over the remaining parameters.
The square root of the diagonal elements will 
give the errors of the corresponding parameter.

\begin{table}
\begin{centering}\begin{tabular}{|c|c|c|c|c|}
\hline & \textbf{Parameters} & $\bf{P\left(k\right)}$ & $\bf{mP(k)}$ & \bf{WL} \tabularnewline
\hline 
\multicolumn{5}{|c|}{ }  \tabularnewline
\hline 
1& total matter density& $\Omega_{m_{0}}h^{2}$ & $\Omega_{m_{0}}h^{2}$ & $\Omega_{m_{0}}h^{2}$ \tabularnewline
\hline 
2 & total baryon density & $\Omega_{b_{0}}h^{2}$& $\Omega_{b_{0}}h^{2}$& $\Omega_{b_{0}}h^{2}$\tabularnewline
\hline 
3 & optical thickness & $\tau$& $\tau$& $\tau$\tabularnewline
\hline 
4 & spectral index & $n_{s}$ & $n_{s}$ & $n_{s}$\tabularnewline
\hline 
5&matter density today& $\Omega_{m_{0}}$& $\Omega_{m_{0}}$& $\Omega_{m_{0}}$\tabularnewline
\hline 
6& equation of state parameter & $w_0$ & $\, $& $w_0$ \tabularnewline
\hline 
7& equation of state parameter & $w_1$ & $\, $ & $w_1$\tabularnewline
\hline 
8 & rms fluctuations & & & $\sigma_8$\tabularnewline
\hline 
\multicolumn{5}{|c|}{ }  \tabularnewline
 \multicolumn{5}{|c|}{\bf{For each redshift bin}}  \tabularnewline
\multicolumn{5}{|c|}{ } \tabularnewline
\hline
9 & growth index & $\gamma\lzr$ or $\{\gamma_0\,,\gamma_a\}$ & & $\gamma\lzr$ or $\gamma_0$\tabularnewline
\hline 
10 & Hubble parameter & & $\log H(z)$ &\tabularnewline
\hline 
11 & Angular diameter distance & & $\log D_{A}(z)$ &\tabularnewline
\hline
12 & Growth factor & & $\log G(z)$ & \tabularnewline
\hline 
13 & z-distortion & & $\log\beta(z)$  &\tabularnewline
\hline
14 & shot noise & $P_{s}$ & $P_{s}$ & $\,$\tabularnewline
\hline 
\end{tabular}\par\end{centering}
\caption{Cosmological parameters for the complete and marginalized ($mP(k)$) galaxy power spectrum 
and weak lensing.\label{tab:cosmological-parameters}}
\end{table}

\subsection{Marginalized galaxy power spectrum}

This approach consists on evaluating the Fisher matrix $F_{ij}$ first for the 
main observables shown in Tab.~\ref{tab:cosmological-parameters} and 
then project into the dark energy parameters, $w_0$, $w_a$ and $\gamma$.
Since we want to propagate the errors to the cosmological parameters above, 
we need to change parameter space. This will be done taking the inverse 
of the Fisher matrix $F_{ij}^{-1}$ and then extracting a submatrix, 
called $F_{mn}^{-1}$ containing only the rows and columns with 
the parameters that depend on $w_0$, $w_a$ and $\gamma$, namely 
$H(z)$, $D_{A}(z)$, $G(z)$ and 
$\beta(z)$.
Then we contract the inverse of the submatrix with the new set of 
parameters; the new Fisher matrix will be given by
\begin{equation}
S_{ij}=\frac{\partial p_m}{\partial q_i}F_{mn}\frac{\partial p_n}{\partial q_i}\,.
\end{equation}
The  square root  of the diagonal elements of the matrix $S_{ij}$ gives the errors 
on the parameters $w_0$, $w_a$ and $\gamma\lzr$. The growth index parameter will be discussed 
 in more detail in the next section.

\subsection{Weak lensing}

Following \cite{schimd, acqua}, the lensing potential is $\Phi_{\rm L} = \Psi +\Phi$, which describes the deviation of light rays. 
As previously mentioned, we limit ourselves to scalar perturbations at linear order in 
the Newtonian gauge Eq.~(\ref{eq:newt-metric}); as we are considering  dark energy only at background level, 
the gravitational potential is simply given by Eq.~(\ref{eq:potential-grav}). As $\Phi=\Psi$ it follows that 
the weak lensing potential is:
\be
k^2\Phi_{\rm L} = 2\frac{3H_0^2\OM}{2a}\Delta_{m}\,.
\ee
The convergence weak lensing power spectrum (which in the linear regime is equal to 
the ellipticity power spectrum) is a linear function of the matter power spectrum convoluted 
with the lensing properties of space. For a $\Lambda$CDM cosmology it can be written as \cite{hujain}
\be
P_{ij}(\ell) = H_0^4\int_{0}^{\infty}{\frac{{\rm d}z}{H(z)}W_{i}\lzr W_{j}\lzr P_{nl}\left[P_{l}\left(\frac{H_0\ell}{r\lzr}, z\right)\right]}
\ee
where the $W_{i}$'s are the window functions, $P_{nl}\left[P_{l}\left(k, z\right)\right]$ is the non linear power spectrum 
at redshift $z$ obtained correcting the linear matter power spectrum $P_{l}\left(k, z\right)$, see \cite{aks} for more details.

The Fisher matrix for weak lensing is given by:
\be
F_{\alpha\beta} = f_{\rm sky}\sum_{\ell} \frac{\left(2\ell+1\right)\Delta\ell}{2}\partial\left(P_{ij}\right)_{,\alpha}C_{jk}^{-1}\partial\left(P_{km}\right)_{,\beta}C_{mi}^{-1}
\ee
where the partial derivatives represent $\partial /\partial\theta_{\alpha}$, the corresponding cosmological parameters $\theta_{\alpha}$ are shown in Tab.~\ref{tab:cosmological-parameters} and
\be
C_{jk}= P_{jk}+\delta_{jk}\frac{\langle \gamma_{\rm int}^{1/2}\rangle}{n_j}
\ee
where $\gamma_{\rm int}$ is the rms intrinsic shear (here we assume $\langle\gamma_{\rm int}^{1/2}\rangle$=0.22 \cite{amara}) 
and $n_{j}$ is the number of galaxies per steradians belonging to the $i$-th bin.

\section{Results}\label{sec:results}

We are now in the position to derive the sensitivity of a typical next-generation survey to the 
dark energy parameters. 

In particular, we consider a survey patterned according to the specification of the Euclid survey, see \cite{euclid}. 
In order to do this we make use of the Fisher matrix formalism for the three different  observables
introduced in the previous section: complete and marginalized galaxy power spectrum and Weak Lensing.

{\bf Galaxy redshift survey specifications}: for the galaxy power spectrum cases we consider a spectroscopic 
survey from $z = 0.5 - 2.1$ divided in equally spaced bins of width $\Delta z = 0.2$ and with a 
covering area of $20000\,{\rm deg}^2$; 
the galaxy number densities in each bin are shown in Tab.~\ref{tab:survey} with an efficiency of 50\%. For all cases we assume that the error on the measured redshift is spectroscopic, $\delta z = 0.001\left(1+z\right)$.

While modeling the redshift survey, we choose two different cases according to 
the galaxy number density \cite{geach, portoamendola}: 
\begin{itemize}
\item optimistic case; this corresponds to the middle column in Tab.~\ref{tab:survey}.
\item realistic case; this corresponds to the last column in Tab.~\ref{tab:survey}.
\end{itemize}

\begin{table}
\begin{centering}\begin{tabular}{|c|c|c|}
\hline 
$\bf{z}$ & $\bf{n_{1}(z)\times 10^{-3}}$ &$\bf{n_{2}(z)\times 10^{-3}}$\tabularnewline
\hline 
\multicolumn{3}{|c|}{ }  \tabularnewline
\hline
$0.5-0.7 $& $4.69$ & $3.56$ \tabularnewline
\hline 
$0.7-0.9 $& $3.33$ & $2.42$ \tabularnewline
\hline 
$0.9-1.1 $& $2.57$ & $1.81$ \tabularnewline
\hline 
$1.1-1.3 $& $2.1$ & $1.44$ \tabularnewline
\hline 
$1.3-1.5 $& $1.52$ & $0.99$ \tabularnewline
\hline 
$1.5-1.7 $& $0.92$ & $0.55$ \tabularnewline
\hline 
$1.7-1.9 $& $0.54$ & $0.29$ \tabularnewline
\hline 
$1.9-2.1 $& $0.31$ & $0.15$ \tabularnewline
\hline 
\end{tabular}\par\end{centering}
\caption{Expected galaxy number density in each redshift bin for the
Euclid survey in units of $\left(h/{\rm Mpc}\right)^3$.
\label{tab:survey}}
\end{table}

\

{\bf Weak lensing survey specifications:} 
for the WL survey we consider a photometric survey characterized by the sky fraction $f_{\rm sky}=1/2$, 
that is, a covering area of $20000{\rm deg}^2$; an overall 
radial distribution $n\lzr=z^2\,{\rm exp}\left[-\left(z/z_0\right)^{1.5}\right]$ with 
$z_{0} = z_{\rm mean}/1.412$ and  mean redshift $z_{\rm mean}=0.9$, the number 
density is $d=35$ galaxies per arcmin$^2$. Moreover, we consider the range $10<\ell<10000$ and we extend our survey 
up to $z_{\rm max}=3$ divided in 5  bins each containing the same number of galaxies. For the non linear 
correction we use the halo fit model by Smith et al. \cite{smithal}. 
We assume the error on the measured redshift is photometric, $\delta z = 0.05\left(1+z\right)$.

\

{\bf Fiducial model}: our fiducial model corresponds to the $\Lambda$CDM WMAP-7yr 
\cite{komatsu}: $\Omega_{m,0}h^2=0.134$, $\Omega_{b}h^2=0.022$, $n_s=0.96$, $\tau=0.085$, 
$h=0.7$, $\Omega_{m,0}=0.275$ and $\Omega_{K}=0$. For the dark energy parameters 
we choose $w_0=-1$ and $w_a=0$.

\vspace{0.2in}

We can now derive the sensitivity of the parameters introduced above. 
More specifically, we consider separately three different cases: 
in Case 1 the growth index has been chosen to be a free parameter but 
independent in each redshift bin; in Case 2 we consider the growth index 
as a free parameter but equal for all the redshift bins; in Case 3 
the growth index depends directly on the equation of state parameters ($w_0$ and $w_a$), 
using Eq.~(\ref{eq:gamma-2f1-wa}) as the analytic solution. 

\

\begin{itemize}
\item {\bf Case 1:}

We consider the growth index $\gamma$ as a free parameter and independent for each redshift bin 
in order to map its variation over time; its value is chosen to be $0.545$. 
The errors are shown in Fig.~\ref{fig:gamma-errors-z} for both 
galaxy power spectrum cases; in Tab.~\ref{tab:gamma-errors-z} are reported their 
$1-\sigma$ errors. Overall the errors on the growth index are of about $0.02$ and $0.03$ for the 
complete and marginalized galaxy power spectrum, respectively. 
In Fig. \ref{fig:wl-anal} the errors for the WL case are shown. It is worth noticing that after $z\sim 1$ the errors 
on $\gamma$ are basically unchanged. 
Furthermore, we also plot the analytic expression for the growth index Eq.~(\ref{eq:gamma-2f1-wa}) 
as a comparison. However, the difference between the full analytic expression for the growth index and 
its asymptotic value $\gamma = 0.545$ is too small to be detected even for a half sky survey like Euclid for 
both the spectroscopic and the photometric cases.

\begin{figure}
\centering
\includegraphics[scale=0.69]{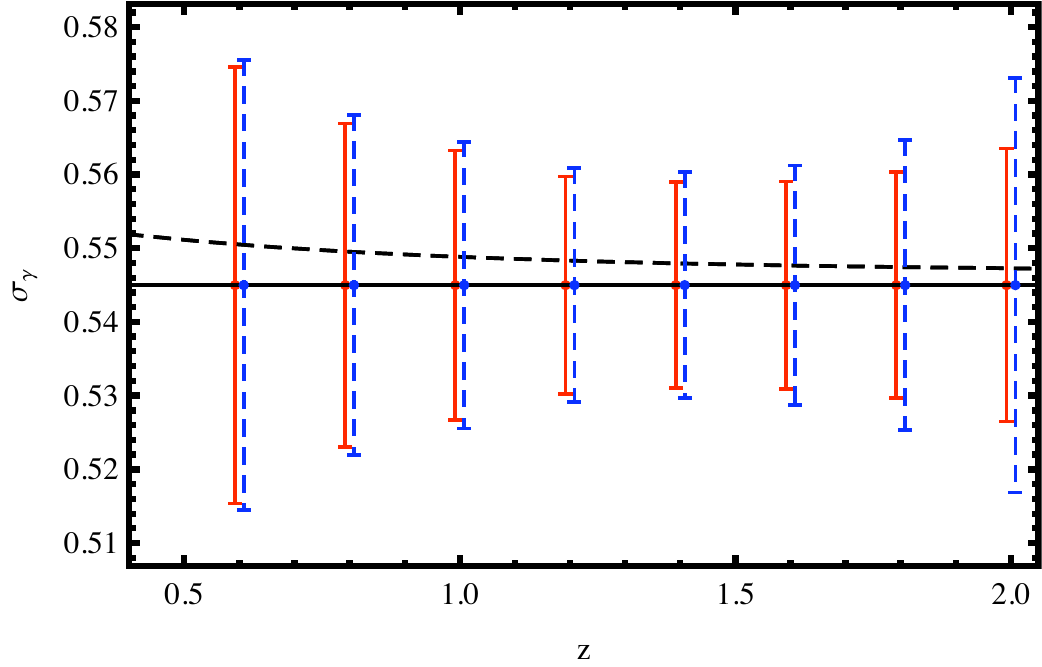}
\hspace{0.1in}
\includegraphics[scale=0.69]{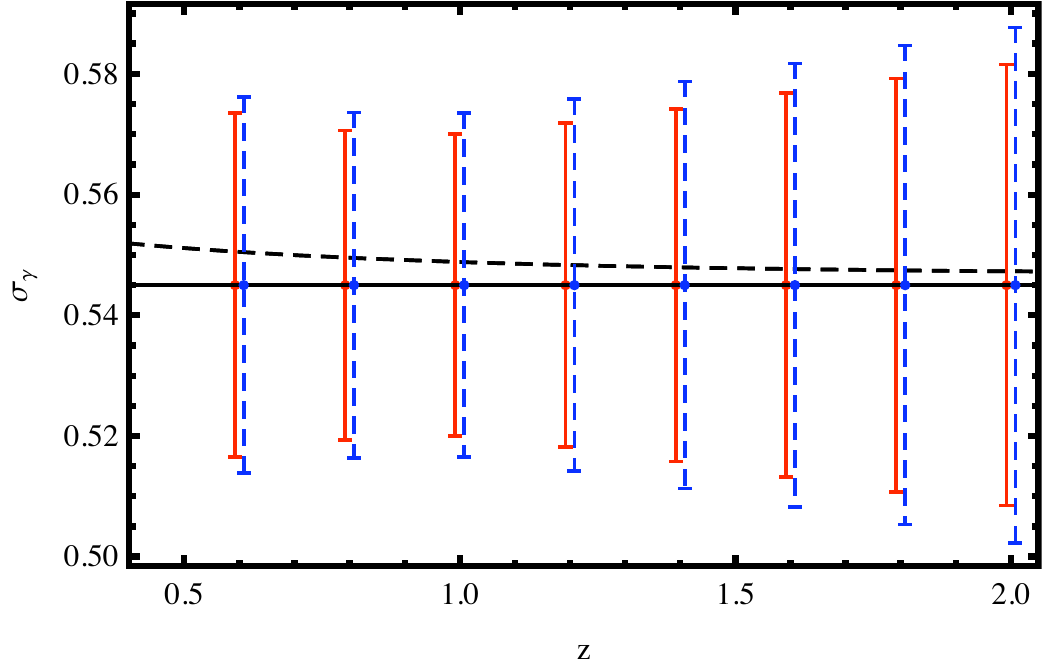}
\caption{\label{fig:gamma-errors-z}Marginalized errors for the growth index 
$\gamma$ at different redshift bins for the complete (left panel) and marginalized (right panel) galaxy power spectrum. 
The blue dashed and the red errors bars refer to the 
optimistic and realistic case, respectively. The black long dashed line is the 
growth index give by the Eq.~(\ref{eq:gamma-2f1-wa}).}
\end{figure}

\begin{figure}
\centering
\includegraphics[scale=0.65]{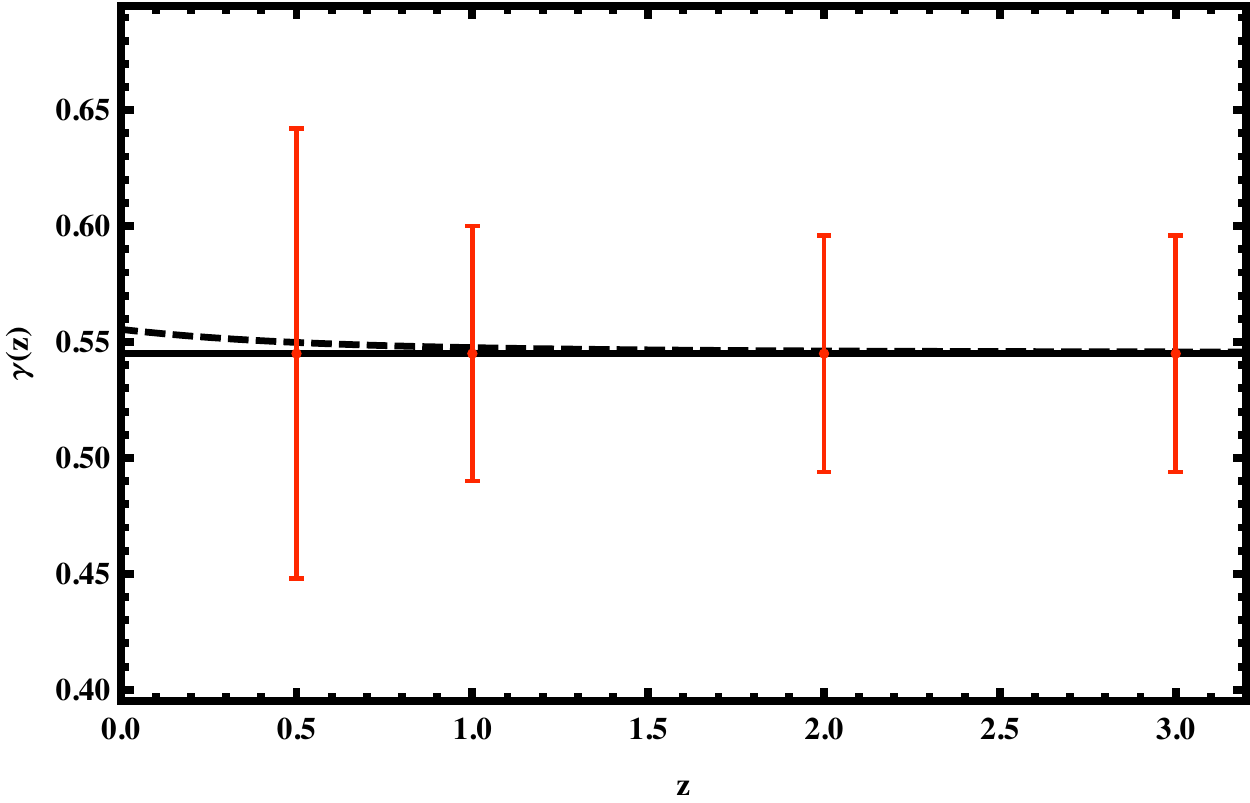}\\
\caption{\label{fig:wl-anal}
Marginalized errors for the growth index $\gamma$ at different redshift bins for the WL convergence spectrum.}
\end{figure}


\begin{table}
\begin{centering}
\begin{tabular}{|c|c|c|c|c|c|c|c|c|}
\hline 
\multicolumn{9}{|c|}{$\bf{P\left(k\right)}$}\tabularnewline
\hline 
\hspace{0.5cm}$z$\hspace{0.5cm}  & $0.6$ & $0.8$ & $1.0$ & $1.2$ & $1.4$ & $1.6$ & $1.8$ & $2.0$ \tabularnewline
\hline 
$\sigma_\gamma^{real}$ & $0.0305$ & $0.0231$ & $0.0194$ & $0.0159$ & $0.0153$ & $0.0162$ & $0.0197$ & $0.0281$  \tabularnewline
\hline 
$\sigma_\gamma^{opt}$ & $0.0296$ & $0.0219$ & $0.0183$ & $0.0147$ & $0.0139$ & $0.0141$ & $0.0153$ & $0.0185$  \tabularnewline
\hline 
\multicolumn{9}{|c|}{}\tabularnewline
\multicolumn{9}{|c|}{$\bf{mP(k)}$}\tabularnewline
\hline 
\hspace{0.5cm}$z$\hspace{0.5cm}  & $0.6$ & $0.8$ & $1.0$ & $1.2$ & $1.4$ & $1.6$ & $1.8$ & $2.0$ \tabularnewline
\hline 
$\sigma_\gamma^{real}$ & $0.0311$ & $0.0286$ & $0.0284$ & $0.0308$ & $0.0337$ & $0.0367$ & $0.0397$ & $0.0427$ \tabularnewline
\hline 
$\sigma_\gamma^{opt}$ & $0.0286$ & $0.0256$ & $0.0250$ & $0.0268$ & $0.0292$ & $0.0318$ & $0.0342$ & $0.0366$ \tabularnewline
\hline 
\end{tabular}
\par\end{centering}
\caption{Here are listed the $1-\sigma$ errors for the growth index $\gamma$ at 
different redshifts for the $P\left(k\right)$ methods.}\label{tab:gamma-errors-z}
\end{table}

\begin{table}
\begin{centering}
\begin{tabular}{|c|c|c|c|c|c|c|c|c|}
\hline 
\multicolumn{5}{|c|}{\bf{WL}}\tabularnewline
\hline 
\hspace{0.5cm}$z$\hspace{0.5cm}  & $0.5$ & $1.0$ & $2.0$ & $3.0$ \tabularnewline
\hline 
$\sigma_\gamma$ & $0.097$ & $0.055$ & $0.051$ & $0.051$  \tabularnewline
\hline 
\end{tabular}
\par\end{centering}
\caption{Here are listed the $1-\sigma$ errors for the growth index $\gamma$ at 
different redshifts for the WL method.}\label{tab:gamma-errors-wl}
\end{table}

\vspace{3mm}

\item {\bf Case 2:}

For this case we chose the growth index to be \cite{portoamendola,wu}
\be\label{gammapara}
\gamma(z) = \gamma_{0}+\gamma_{a}(1-a) = \gamma_{0}+\gamma_{a}\frac{z}{1+z}
\ee
in order to compare the alternative theories introduced in the previous sections. 
Our reference cosmology is always $\Lambda$CDM with $\gamma_0=0.55620$ and $\gamma_a=0.0182537$. 
We choose these values according to the Taylor's expansion of Eq.~(\ref{eq:gamma-2f1-wa}) 
for the $\Lambda$CDM cosmology. In Fig.~\ref{fig:3-params-all}  the $1-\sigma$ confidence regions for the 
parameters $w_0-w_a$ (left panel) and $\gamma_0-\gamma_a$ (right panel) for the galaxy power spectrum cases are shown.

For the WL case we found $\gamma_a$ to be degenerate with $\gamma_0$ indicating 
probably that a weak lensing experiment is insensible to the variation of the growth index  with redshift; 
this can be also seen in Fig.~\ref{fig:wl-anal} where the errors of the growth index are basically the same 
for all the redshift bins above $z\simeq 1$. In Tab.~\ref{tab:errors-params} we report the errors for the parameters 
introduced above for all the three surveys.

We reach a sensitivity of about $\sigma_{w_0} = 0.018$, $\sigma_{w_0} = 0.068$ and $\sigma_{w_0} = 0.122$ for the
galaxy power spectrum cases and WL survey, respectively. These errors are sufficiently small to rule out independently 
most of the models we considered in this paper, see Fig.~\ref{fig:3-params-all} where we also plot the expected values of the 
equation of state parameters $w_0-w_a$ and the growth index $\gamma_0-\gamma_a$ for the alternative cosmological models: 
quintessence model (yellow box), DGP (blue diamond), LTB (black triangle) and $f(R)$ model (brown inverted triangle). 
Only a measurement of the equation of state parameter is able to rule out wCDM, DGP and LTB model; however, $f(R)$ is almost 
indistinguishable from the reference cosmology and none of the surveys assumed here are able to rule it out.
Fortunately it is also possible to measure the growth index $\gamma_0$; in this case we reach a sensitivity of 
about $0.02$, $0.092$ and $0.075$ for  the galaxy power spectrum cases and WL survey, respectively. On the other hand,
measuring only $\gamma_0-\gamma_a$, the model that could be clearly ruled out is DGP, while LTB could still be a viable model. 
However, since we are dealing with two different surveys we can consider measuring one set of parameters for each experiment; 
for example, we can measure $w_0-w_a$ with the WL survey and $\gamma_0-\gamma_a$ with a $P(k)$ experiment. In this case 
the models considered in this work would be excluded by at least at the $2\sigma$ level. With respect to $f(R)$ models, note that
the Taylor expansion of $\gamma(a)$ does not work appropriately in the whole range of redshifts explored by the survey. Therefore, 
we have left this class of models out of Fig.~\ref{fig:3-params-all}. However, inspecting Fig.~\ref{fig:gamma-error} one can easily
conclude that models $f(R)$ would also be ruled out by the measurements of a Euclid-like survey at the many $\sigma$ level.

\begin{figure}
\centering
\includegraphics[scale=0.55]{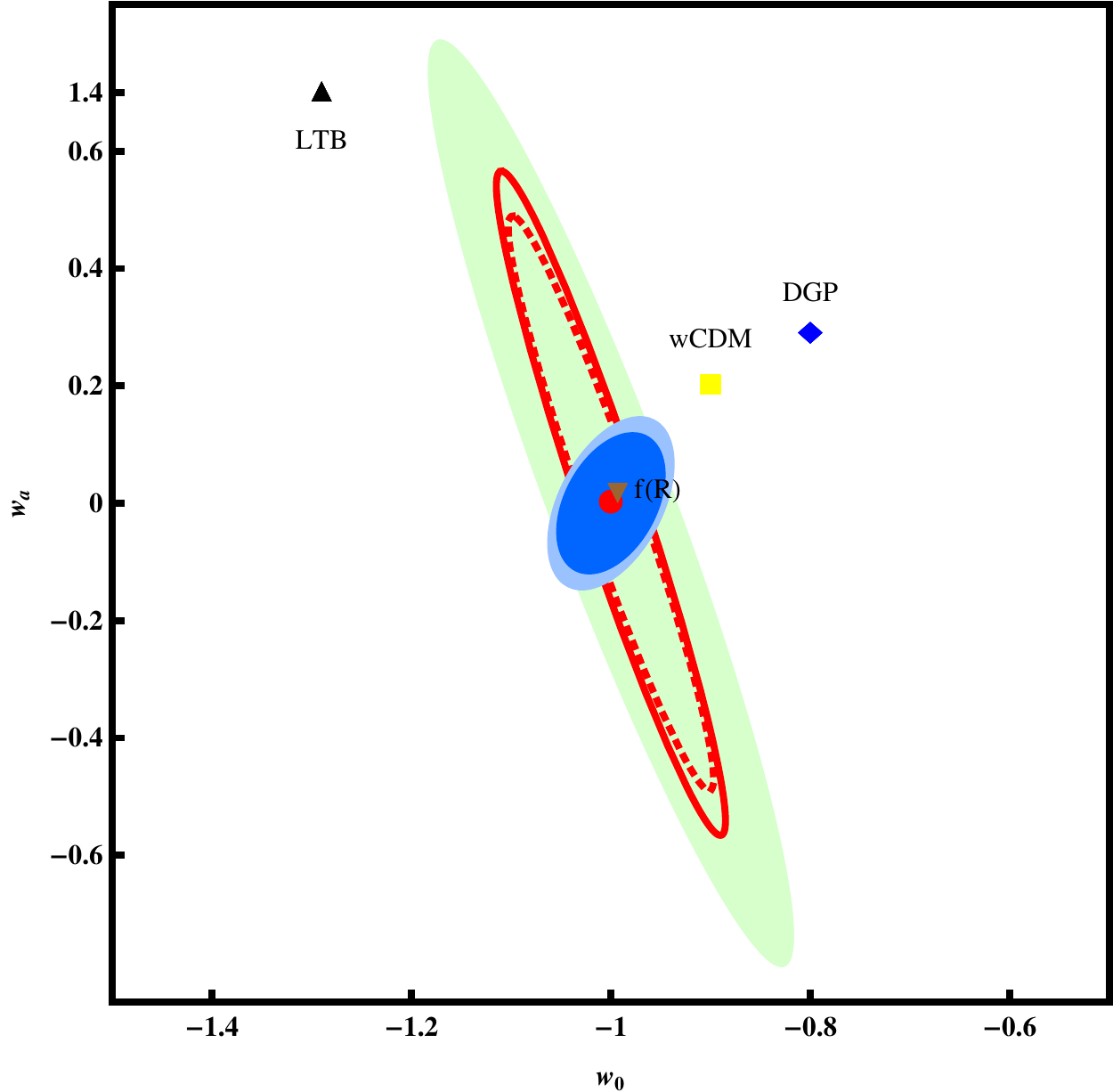}
\hspace{0.1in}
\includegraphics[scale=0.55]{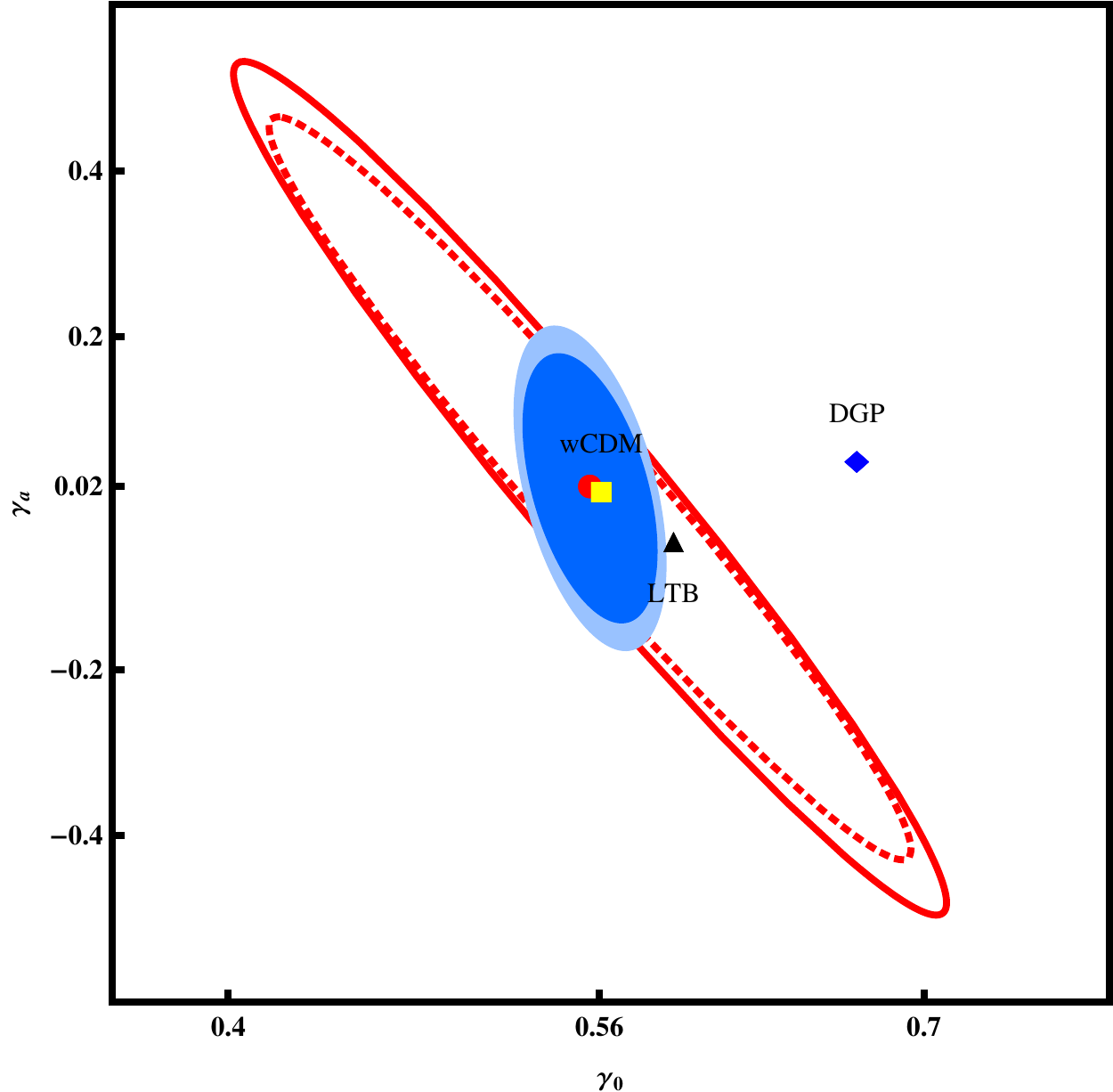}\\
\caption{\label{fig:3-params-all}
Confidence level for the $w_0$, $w_a$ (left panel) and $\gamma_0$ , $\gamma_a$ (right panel). 
The blue and light blue areas are for the galaxy power spectrum case, optimistic and realistic case respectively. The red dashed and solid lines are for the $mP(k)$ case, optimistic and realistic case respectively. The light green shaded area is for the WL case. These two plots correspond to case 2 of our analysis, taking $\gamma= \gamma_0+\gamma_a(1-a)$. Note that the vertical axis has been broken on the left panel plot to include the point corresponding to the LTB model. The points corresponding to the different models are the $w_0,w_a$ and $\gamma_0,\gamma_a$ pairs calculated for each model.}
\end{figure}

\begin{table}
\begin{centering}
\begin{tabular}{|c|c|c|c|c|c|}
\hline 
& \multicolumn{2}{|c|}{$\bf{P\left(k\right)}$}& \multicolumn{2}{|c|}{$\bf{mP(k)}$}& \bf{WL}\tabularnewline
\hline
 & real. & opt. & real. & opt. &  \tabularnewline
\hline
$\sigma_{w_{0}}$ & $0.021$ & $0.018$ & $0.076$ & $0.068$ & $0.122$  \tabularnewline
\hline
$\sigma_{w_{a}}$ &$0.051$ & $0.041$ & $0.375$ & $0.324$ & $0.524$ \tabularnewline
\hline
$\sigma_{\gamma_{0}}$ & $0.022$ & $0.020$  & $0.102$ & $0.092$ & $0.075$\tabularnewline
\hline
$\sigma_{\gamma_{a}}$ & $0.120$ & $0.116$ & $0.339$ & $0.296$ & $\,$\tabularnewline
\hline
\end{tabular}
\par\end{centering}
\caption{Here are listed the $1-\sigma$ errors for $w_0$, $w_a$, $\gamma_0$ 
and $\gamma_a$ for the $P\left(k\right)$, $mP(k)$ and WL cases.}\label{tab:errors-params}
\end{table}

\item {\bf Case 3:}

In this case, the growth index $\gamma$ depends on the cosmological parameters $w_0-w_a$ according to 
Eq.~(\ref{eq:gamma-2f1-wa}); the derivatives of the growth factor are given by:
\be
\frac{\partial\log G}{\partial q_i}= -\int{\left[\frac{\partial\gamma}{\partial q_i}\log\Om_{m}\lzr + \gamma\frac{\partial\log\Om\lzr}{\partial q_i}\right]}\Om\lzr^\gamma\frac{{\rm d}z}{1+z}
\ee 
which has an extra term $\partial\gamma/\partial q_i$, our new set of parameters becomes now $q_i=\{w_0,w_a\}$. 
In Fig.~\ref{fig:all-anal} we plot the $1-\sigma$ confidence regions for $w_0-w_a$ for all the three experiments; 
in the same figures are also plotted the values of the equation of state parameters 
$w_0-w_a$ for the alternative cosmological models: quintessence model (yellow box), DGP (blue diamond) and 
LTB model (black triangle)\footnote{We have not plotted the $f_S\left(R\right)$ model here because the Ansatz for $\gamma(a)$
is not valid in the whole range of redshifts.}. 

However, in order to be able to include in the same graphic the other cosmological models we had to multiply 
the confidence regions by a factor of $5$ for the galaxy power spectrum cases and by a factor 
of $3$ the WL case; in the right panel of Fig.~\ref{fig:all-anal} we zoomed over the confidence region to preserve 
the proportions of the errors obtained with the three different methods. 
Giving a direct dependence to the growth index $\gamma$ on the equation of state parameters the 
errors on  $w_0-w_a$  are reduced by a factor $4$ for the $P(k)$ case, a factor $10$ for the $mP(k)$ and 
of about a factor $2$ for WL. In this case we reach an extreme sensitivity of about $0.5-1.0 \%$ for $w_0$ and of about 
$2-3 \%$ for $w_a$, however still not sufficiently good to rule out completely the $f(R)$ model studied.

\end{itemize}

\begin{figure}
\centering
\includegraphics[scale=0.55]{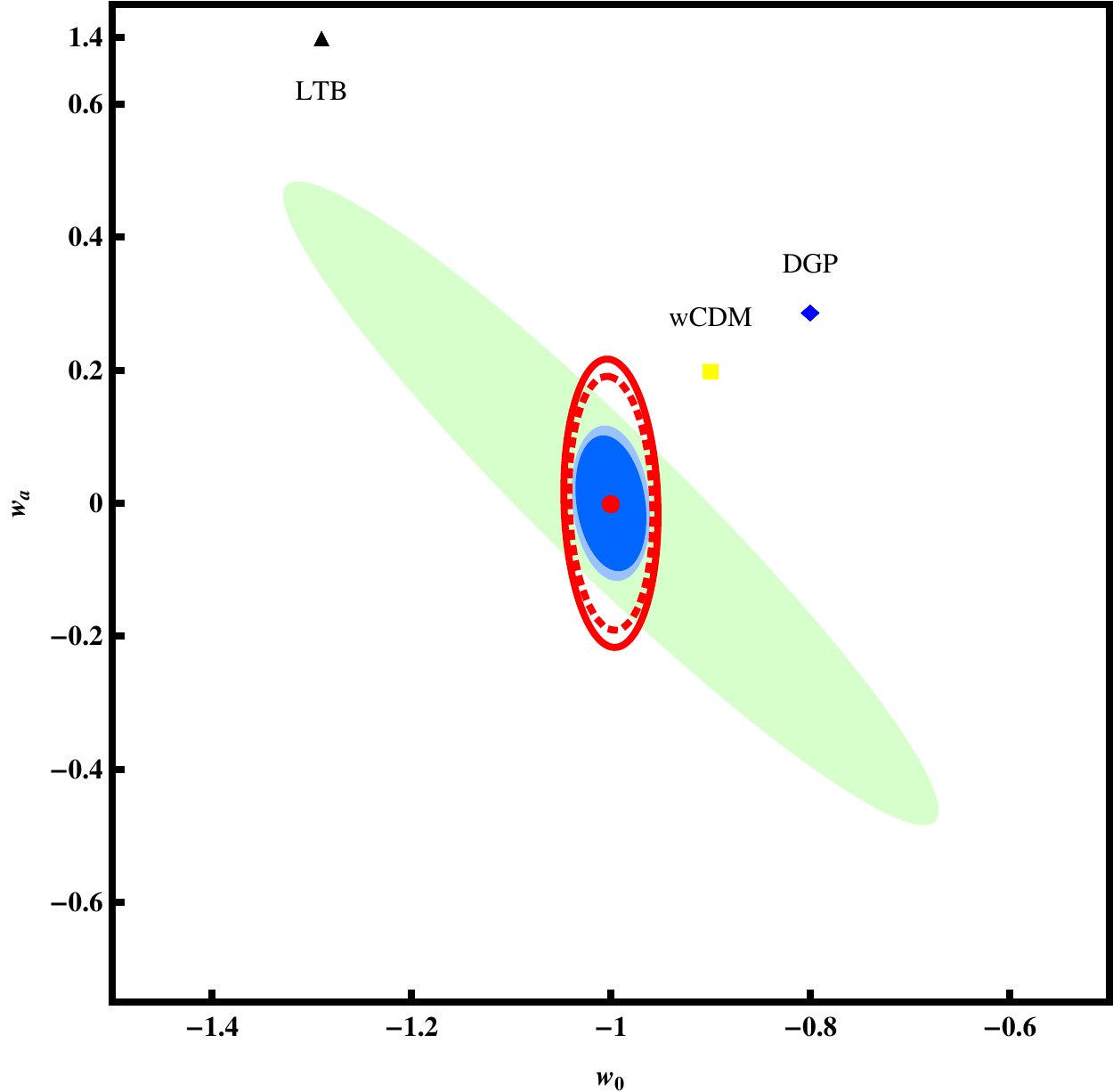} 
\hspace{0.1in}
\includegraphics[scale=0.55]{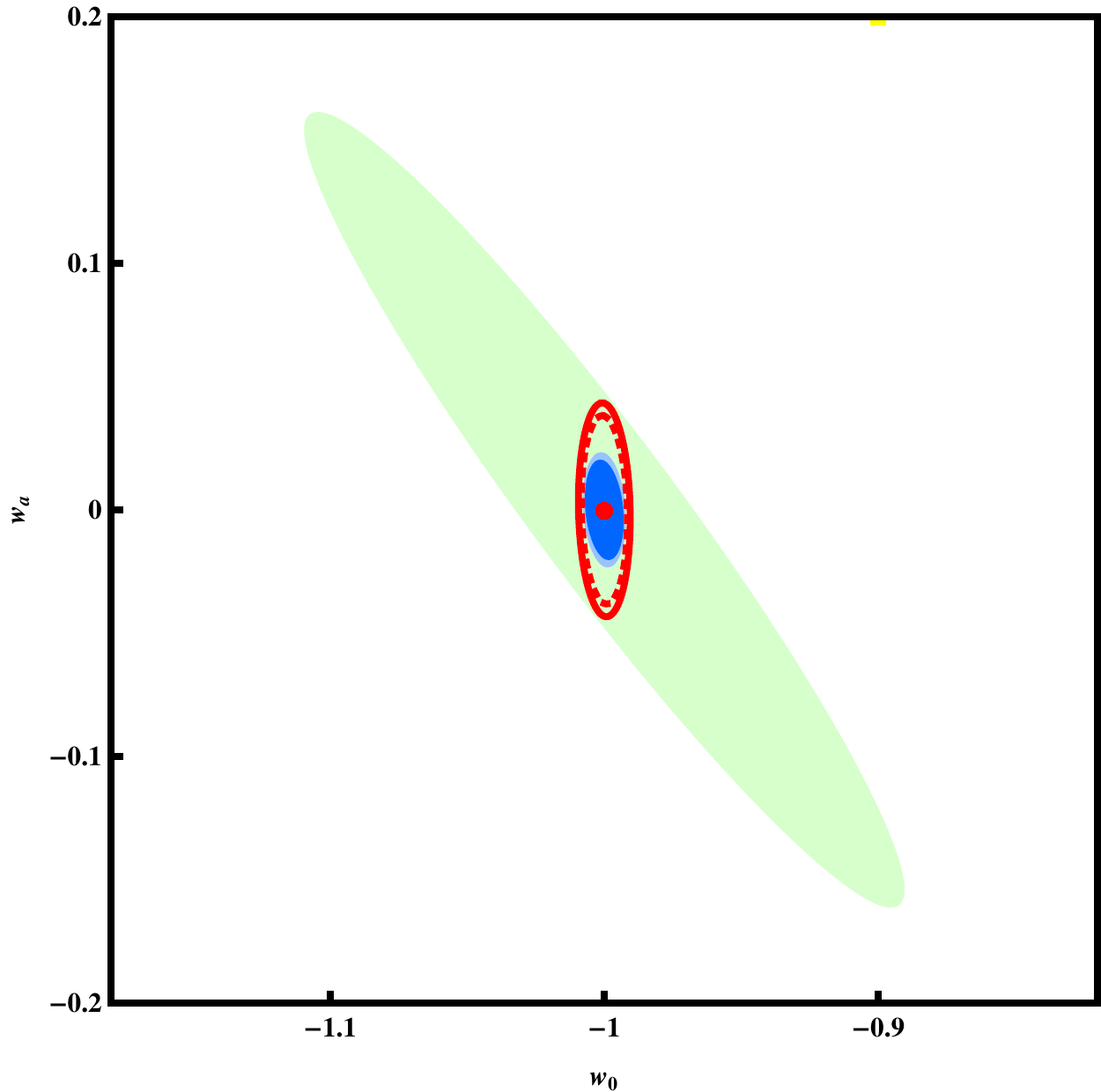} 
\caption{Confidence level for $w_0$ and $w_a$ for all the three cases with the growth index 
given by Eq.~(\ref{eq:gamma-2f1-wa}). The red solid line and blue dashed line refer to the optimistic 
and realistic case, respectively. Note that the vertical axis has been broken on the left panel plot to include the point corresponding to the LTB model. Note that the right figure contains a scaled version of the left figure. For details see the text. The points corresponding to the different models are the $w_0,w_a$ and $\gamma_0,\gamma_a$ pairs calculated for each model.
}
\label{fig:all-anal} 
\end{figure}

\begin{table}
\begin{centering}
\begin{tabular}{|c|c|c|c|c|c|}
\hline 
& \multicolumn{2}{|c|}{$\bf{P\left(k\right)}$}& \multicolumn{2}{|c|}{$\bf{mP(k)}$}& \bf{WL}\tabularnewline
\hline
 & real. & opt. & real. & opt. &  \tabularnewline
\hline
$\sigma_{w_{0}}$ & $0.0052$ & $0.0047$ & $0.0063$ & $0.0056$ & $0.072$  \tabularnewline
\hline
$\sigma_{w_{a}}$ &$0.0155$ & $0.0135$ & $0.0281$ & $0.025$ & $0.106$ \tabularnewline
\hline
\end{tabular}
\par\end{centering}
\caption{Here are listed the $1-\sigma$ errors for $w_0$ and $w_a$  
for the $P\left(k\right)$, $mP(k)$ and WL cases.}\label{tab:errors-params-anal}
\end{table}

\section{Conclusions}\label{s:con}

Over the last decade there has been a plethora of proposals to account for the observed acceleration of the universe. All these proposals fall into four main categories: i) the inclusion of some extra field (scalar, vector or tensor), coupled or not to the rest of matter, like in quintessence, chameleon, vector dark energy or massive gravity; ii) the extension of GR by inclusion of higher order terms in the Einsten-Hilbert action, like $f(R)$ theories, Gauss-Bonnet terms, etc.; iii) the modification of gravity on large scales by introduction of extra dimensions, like in Dvali-Gabadadze-Porrati model, Kaluza-Klein gravity, etc.; iv) the reinterpretation in terms of a nontrivial spatial geometry, like in large-void inhomogeneous LTB models.

All of these proposals have very specific predictions for the background evolution of cosmological space-times, and most of them can be well fitted to the present observations, with just a few phenomenological parameters: the equation of state, the speed of sound, the coupling between DM and DE, bulk viscosity, etc. However, in order to discriminate between the different alternatives it has been realized that one has to go beyond the background evolution and start to consider also the theory of linear cosmological perturbations and parametrize their evolution in terms of the growth function and growth index, as well as the shift parameter.

\begin{figure}
\centering
\includegraphics[scale=1.0]{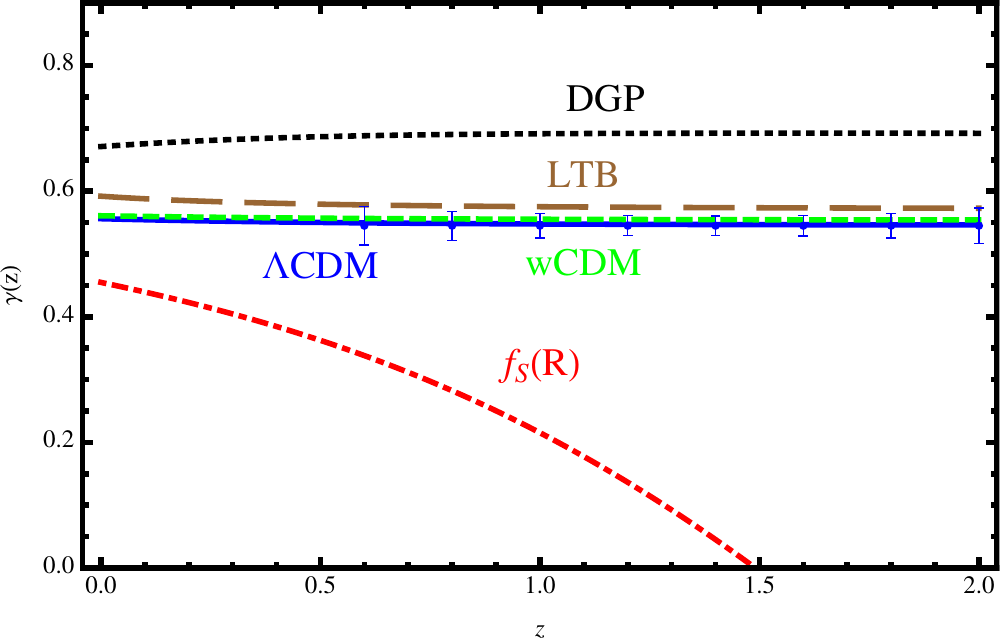}
\caption{The redshift evolution of the growth index compared with the expected errors from a Euclid-like survey. Note that DGP and the studied $f(R)$ models can in principle be ruled out at many sigma.}
\label{fig:gamma-error}
\end{figure}

In this paper we provide exact solutions to the cosmological matter perturbation equation in a homogeneous FLRW universe with a vacuum energy that can be parametrized with a variable equation of state parameter $w(a)=w_0+w_a(1-a)$. We compute the growth index $\gamma=\log f(a)/\log\Om_m(a)$, and its redshift dependence, using the exact solutions in terms of Legendre polynomials and show that it can be parametrized as $\gamma(a)=\gamma_0+\gamma_a(1-a)$ for most cases, see Fig.~\ref{fig:gamma-ansatz}. We then compare four different types of dark energy models: $w\Lambda$CDM, DGP, $f(R)$ and a LTB-large-void model, which have very different behaviors at $z\gsim1$, see Fig.~\ref{fig:gamma-error}. This allows us to study the possibility to differentiate between various alternatives using full sky deep surveys like Euclid, which will measure both photometric and spectroscopic redshifts for several million galaxies up to redshift $z=2$. We do a Fisher matrix analysis for the prospects of differentiating among the different DE models in terms of the growth index, taken as a given function of redshift or with a principal component analysis, with a value for each redshift bin, see also \cite{camera1, camera2} for a similar analysis for DGP and $f(R)$ models. We use as observables the complete and marginalized power spectrum of galaxies $P(k)$  and the Weak Lensing (WL) power spectrum. We find that using $P(k)$ one can reach (2\%, 5\%) errors in $(w_0, w_a)$ and (4\%, 12\%) errors in $(\gamma_0, \gamma_a)$, while using WL we get errors at least twice as large. These estimates allow us to differentiate easily between DGP, $f(R)$ Starobinsky model and $\Lambda$CDM, see Fig.~\ref{fig:gamma-error}, while it would be more difficult to distinguish the latter from a variable wCDM or LTB models using only the growth index.

\section*{Acknowledgments} 

We thank Luca Amendola and Martin Kunz, as coordinators of the Theory WG of the Euclid consortium, 
for a careful reading of the manuscript and useful suggestions.
ABB thanks the Spanish Education Ministry (ME) for support within the FPU grant 
program with ref. number AP2008-02679. 
She also thanks A.~A.~Starobinsky for valuable discussions on the growth index for $f(R)$ theories.
DS acknowledges support from the JAEDoc program with ref. JAEDoc074.
JGB thanks the Institute of Theoretical Physics in Geneva for hospitality during the sabbatical 
year 2009-10, where this work was initiated. We also acknowledge financial support from the Madrid Regional Government 
(CAM) under the program HEPHACOS P-ESP-00346, and MICINN under grant  AYA2009-13936-C06-06. 
We all participate in the Consolider-Ingenio 2010 PAU (CSD2007-00060), 
as well as in the European Union Marie Curie Network 
``UniverseNet" under contract MRTN-CT-2006-035863.

\section*{Appendices}

\appendix
\section{Small scale approximation}\label{app:fullsolution}
In this appendix we present the small scale solution found to the perturbation equation for the matter density contrast. The master equation for the perturbations in the small scale regime (i.e.: large wave number $k$) \eqref{eq:mastereq} can be solved by changing the independent variable from time to the scale factor,
\be
a^2\delta''(a) + (3-\eps(a))\,a\,\delta'(a) - \frac{3}{2}\Om_m(a)\,\delta(a) = 0\,.
\label{eq_smallscales}
\ee
where $\epsilon$ is given by
\be
\epsilon(a)= -\frac{aH'}{H} = \frac{3}{2}\Big[1+w(a)\Big(1-\Om_m(a)\Big)\Big]
\ee
Furthermore, by noting that the density contrast for pure matter grows like $\delta(a)\propto a$,
one can write the general function 
\be
\delta(a) = a\cdot G(a)\,,
\ee
with which the master equation becomes
\be
a^2 G''(a) + \Big(5 - \eps(a)\Big)\,a\,G'(a) + \Big(3 - \eps(a) - \frac{3}{2}\Om_m(a)\Big)\,G(a) = 0\,.
\label{eq:master-g}
\ee
Now we make a change of variables,
\be
u = -\frac{\ODE}{\OM}\,a^{-3w}\,, \hspace{5mm} \Longrightarrow \hspace{5mm}
a \frac{d}{da} = -3w\,u \frac{d}{du}\,, \hspace{5mm} \Om_{m}(a) = \frac{1}{1-u} \,,
\ee
and thus the master equation becomes
\be
u(1-u) G''(u) + \left[1 - \frac{5}{6w} - \Big(\frac{3}{2} - \frac{5}{6w}\Big)u \right] G'(u) 
- \frac{1-w}{6w^2}\,G(u) = 0\,,
\label{eq:master-g-u}
\ee
which has the form of a Hypergeometric equation, see Ref.\cite{AS},  with constant
coefficients $\alpha=(w-1)/2w, \beta=-1/3w, \gamma=1-5/6w$, and thus the exact solutions are 
written in terms of two independent constants, $C_1$ and $C_2$,
\bea
\delta(a) &\!=\!& C_1\,a\cdot {}_2\!F_1\left(\frac{w-1}{2w},\frac{-1}{3w},1-\frac{5}{6w}; 
- \frac{\ODE}{\OM}\,a^{-3w}\right) \\
&\!+\!& C_2\,a^{-3/2}\cdot {}_2\!F_1\left(\frac{1}{2w},\frac{1}{2}+\frac{1}{3w},1+\frac{5}{6w}; 
- \frac{\ODE}{\OM}\,a^{-3w}\right)\,. \nonumber
\eea
The first term corresponds to the growing mode solution and the second one to the
decaying mode. When describing late time solutions we will always take the
growing mode solution; furthermore, the integration constant $C_1$ is not a problem, as in this 
work we are interested in the evolution of the growth rate $f(a)$ which is the ratio of 
the matter density and its derivative. 

There are a number of way that one can play with the solution above in order to simplify the 
Hypergeometric function and even to drop the Hypergeometric functions. 
Here we use one relation which seems to be the easiest among all the others; 
we can notice that the third coefficient $\gamma$ of the Hypergeometric function can 
be written as:
\be
\gamma = \frac{1}{2}+\alpha+\beta
\ee
being $\alpha$ and $\beta$ the first and second coefficient of the Hypergeometric function. 
In this case the solution for the matter density contrast can be written in terms of 
Legendre polynomials. 
However, being our goal to find an exact solution for the growth rate and growth index, we find 
easier to first evaluate these terms using the Hypergeometric functions and then to simplify the 
result. 
The growth rate is defined as
\be
f(a) = \frac{a\delta'(a)}{\delta(a)} = 1+a\frac{\alpha\beta}{\gamma}\frac{\Om'_{m}(a)}{\Om^2_{m}(a)}\frac{\hyper\Big[\alpha+1, \beta+1, \gamma+1, 1-\Om^{-1}_m(a)\Big]}{\hyper\Big[\alpha, \beta, \gamma, 1-\Om^{-1}_m(a)\Big]}\,,
\ee 
using Eqs.~($15.4.12$) and ($15.4.21$) of \cite{AS} we have:
\be
f(a) = 1+6w\alpha\beta\sqrt{1-\Om_{m}(a)}\frac{P_{\beta-\alpha-\frac{1}{2}}^{-\beta-\alpha-\frac{1}{2}}\left[1/\sqrt{\Om_{m}(a)}\right]}{P_{\alpha-\beta-\frac{1}{2}}^{\frac{1}{2}-\beta-\alpha}\left[1/\sqrt{\Om_{m}(a)}\right]}\,
\ee
being $P_n^m\left(x\right)$ the Legendre polynomial. 
The last equation can be further simplified making using of the recurrence relations of the 
Legendre polynomials, see Eqs.(8.5.3) and (8.5.5) of Ref.~\cite{AS}, then we find
\be
f(a)\ =\ \Om^{1/2}_{m}(a)\ \frac{P_{1/6w}^{5/6w}\left[\Om^{-1/2}_{m}(a)\right]}{P_{-1/6w}^{5/6w}
\left[\Om^{-1/2}_{m}(a)\right]}
\ee
We can then express the growth index $\gamma(a)$ in terms of the Legendre polynomials, and we find:
\be
\gamma(a) = \frac{1}{2}+\frac{1}{\ln\Om_{m}(a)}\ln\left[\frac{P_{1/6w}^{5/6w}\left[\Om^{-1/2}_{m}(a)\right]}{P_{-1/6w}^{5/6w}\left[\Om^{-1/2}_{m}(a)\right]}\right]\,.
\ee

Note that the growth index parameter is close to $0.5\pm0.1$, depending on the range of values of the equation of state parameter and the matter content of the universe.


\subsection{Varying equation of state parameter}

We can extend our discussion also to a varying dark energy equation of state 
parameter $w(a)$. It is fair to be said that in the last case there is no 
exact analytic solution for the matter density contrast as it was for the case in 
which $w$ is constant; the main problem here is that there is no a direct 
transformation between the scale factor $a$ and the new variable $u$. 

Here we assume the equation of state parameter to be:
\be
w(a) = w_0+w_a\,(1-a)\,,
\ee
for which the matter density parameter can be integrated
\be
\Om_m(a) = \left(1 + \frac{\ODE}{\OM}a^{-3(w_0+w_a)}\,e^{3w_a(a-1)}\right)^{-1} \,,
\ee

The master equation for a varying equation of state parameter still looks the same as Eq.~(\ref{eq:master-g}), 
except that now $w$ is a function of the scale factor.
In order to integrate out  we use again the variable 
\be
u=-\frac{1-\OM}{\OM}a^{-3\hat{w}(a)}\hspace{0.2in} {\rm with}\hspace{0.2in}
\hat{w}(a)=\frac{1}{\ln a}\int_{1}^{a}{\frac{w\left(a'\right)}{a'}{\rm d}a'}\,.
\ee
However, we can make the approximation that the $w(a)$ is slowly varying with time and integrate 
out the master equation. 
The approximation is not so rude. To see this, we need to have another look at the master equation 
for matter perturbations.

In this case the dark matter density contrast reads
\be
\delta(a) = a\hyper
\left[\alpha,\beta, \frac{1}{2}+\alpha+\beta; 1-\Om_m^{-1}(a)\right]
\label{eq:delta-anal-w-a}
\ee
with parameters
\bea
\alpha &\!=\!&\frac{w(a)-1}{2w(a)}-\frac{w_a a}{6w^2(a)}\,, \\[1mm]
\beta &\!=\!&\frac{-1}{3w(a)}+\frac{w_a a}{6w^2(a)}\,,
\eea
which gives a density growth function
\be\label{faapprox}
f(a)\ =\ \Om^{1/2}_{m}(a)\ \frac{P_{1/6w(a)}^{5/6w(a) + w_a a/6w^2(a)}
\left[\Om^{-1/2}_{m}(a)\right]}{P_{-1/6w(a)}^{5/6w(a) + w_a a/6w^2(a)}
\left[\Om^{-1/2}_{m}(a)\right]}\,.
\ee
Comparison of the numerical solution for the density growth function 
with the approximate expression~(\ref{faapprox}) shows accordance 
within less than 0.1\% for a very wide range of values of $-2<w_a <2$.

\end{document}